\documentclass[12pt,preprint,xdvi]{aastex}

\usepackage{natbib}
\usepackage{graphicx}

\newcommand{\be}{\begin{equation}}
\newcommand{\ee}{\end{equation}}
\newcommand\eq{Equation}
\newcommand\eqs{Equations}
\newcommand\fig{Figure}
\newcommand\figs{Figures}

\newcommand{\avg}[1]{{\langle{#1}\rangle}}
\newcommand{\half}{\hbox{${1\over2}$}}

\newcommand{\nhat}{{\bf \hat{n}}}
\newcommand{\xvec}{{\bf x}}
\newcommand{\bvec}{{\bf B}}
\newcommand{\reg}{{\cal R}}
\newcommand{\regs}{{\cal S}}
\newcommand{\bvol}{{\cal B}}
\newcommand{\mhvol}{{\cal G}_{MH}}

\begin{document}

\title{Effects of partitioning and extrapolation on 
the connectivity of potential magnetic fields}

\author{D.W. Longcope,$^1$ G. Barnes,$^2$ and C. Beveridge$^1$}
\affil{1. Department of Physics, Montana State University\\
  Bozeman, Montana 59717\\
2. Colorado Research Associates Division, NorthWest Research Associates\\ 
  3380 Mitchell Ln., Boulder CO 80301}


\begin{abstract}
Coronal magnetic field may be characterized by how its field lines
interconnect regions of opposing photospheric flux -- its
connectivity.  Connectivity can be quantified as the net flux
connecting pairs of opposing regions, once such regions are
identified.  One existing algorithm will partition a typical active
region into a number of unipolar regions ranging from a few dozen to a
few hundred, depending on algorithmic parameters.  This work explores
how the properties of the partitions depend on some algorithmic
parameters, and how connectivity depends on the coarseness of
partitioning for one particular active region magnetogram.  We find
the number of connections among them scales with the number of regions
even as the number of possible connections scales with its square.

There are several methods of generating a coronal field, even a
potential field.  The field may be computed inside conducting
boundaries or over an infinite half-space.  For computation of
connectivity, the unipolar regions may be replaced by point sources or
the exact magnetogram may be used as a lower boundary condition.  Our
investigation shows that the connectivities from these various fields
differ only slightly -- no more than 15\%.  The greatest difference is
between fields within conducting walls and those in the half-space.
Their connectivities grow more different as finer partitioning creates
more source regions.  This also gives a quantitative means of
establishing how far away conducting boundaries must be placed in
order not to significantly affect the extrapolation.  For identical
outer boundaries, the use of point sources instead of the exact
magnetogram makes a smaller difference in connectivity: typically 6\%
independent of the number of source regions.
\end{abstract}

\date{Draft: \today}

\section{Introduction}

According to prevailing understanding, coronal activity on the Sun
involves 
energy stored in its magnetic field.  The coronal field is stressed as
the photospheric regions to which it is anchored slowly evolve.  In
order to build a quantitative model based on this insight it is
essential to quantify how the coronal field links these 
photospheric regions -- its connectivity.

Several previous studies have focused on local properties 
of magnetic connectivity as characterized by the point-wise
mapping of positive footpoints to negative footpoints
\citep{Seehafer1986,Low1987,Inverarity1997,Titov2002,Titov2003}.  
This mapping is discontinuous at coronal current sheets (tangential
discontinuities) where reconnection and energy dissipation are
particularly rapid.  Regions where the mapping is extremely distorted,
called quasi-separatrix layers, may play an equally significant role 
in these processes as well
\citep{Longcope1994c,Priest1995,Demoulin1996,Demoulin1997}

While reconnection is a local process, energy storage 
is global, so its study requires a global
characterization of connectivity.  For example, when two active
regions interact magnetically, there is a change in connectivity
whereby new field lines are forged to connect the positive polarity of
one to the negative polarity of the other 
\citep{Sweet1958b,Longcope2005}.
One method of quantifying the global connectivity is to 
group photospheric footpoints into a number of unipolar regions.  This
is a tacit step in, for example, the above reference to the ``positive
polarity of the active region''.
Coronal field lines are then categorized by the
regions to which their positive and negative footpoints belong.

Global connectivity may be used to quantify the energy stored as
coronal field evolves.  
Barring magnetic reconnection, emergence 
or submergence, the total flux in each connection is
preserved even as the photospheric regions move and deform.  The
preservation of connection fluxes constitutes a set of constraints
which can be used to place a lower bound on the coronal magnetic free
energy \citep{Longcope2001b}.

The process of grouping photospheric footpoints into unipolar regions,
called {\em partitioning}, is natural in certain idealized models
\citep{Sweet1958b,Gorbachev1988,Brown1999b}
or in the flux elements of the quiet Sun 
\citep{Schrijver1997,Hagenaar2001,Parnell2002,Welsch2003,%
Close2004,DeForest2007}.  
Magnetograms of real active regions, on the 
other hand, show photospheric field distributed in a complex pattern
whose reduction to regions is less straightforward.  One algorithm
developed by \citet{Barnes2005} uses the vertical field in a
gradient-based tessellation method \citep{Hagenaar1997}.  
This breaks an active region into
anywhere from dozens to hundreds of unipolar photospheric regions,
depending on algorithmic parameters.  Significantly, the regions
identified by the algorithm track inferred photospheric motions
\citep{Longcope2007b}, so their connectivity can be used to bound
coronal energy.  Larger numbers of regions will lead to a larger
number of constraints, and therefore a more restrictive lower bound on
free energy \citep{Longcope2001b,Longcope2007c}.

The connectivity between photospheric regions depends entirely on the
coronal magnetic field anchored to it.  Since high spatial resolution 
measurements are made only at the lowest level of the atmosphere, such
as the photosphere, it is necessary to extrapolate these data into the
corona before connectivity can be determined.  There are numerous
methods for performing this extrapolation 
\citep[see][for a review]{McClymont1997} and each one
will produce a different connectivity.  The most sophisticated class
of methods, the 
non-linear force-free field (NLFFF) models, includes at least a half
dozen variants, many of which 
have recently been inter-compared in a series of
investigations \citep{Schrijver2006,nlfff2,nlfff3}.  The connectivity was used as a
basis of comparison in one of these investigation \citep{nlfff2}, and it was 
found that the different NLFF fields
produced by these methods each induced a different
connectivity.  Indeed, the differences in connectivity tended to be large 
even when other metrics showed reasonable agreement between an extrapolation 
and the model field.  It is therefore essential
to understand how much the connectivity might vary under different
extrapolations.  This is the objective of the present study.

At the opposite extreme to NLFFF models is the potential
field extrapolation, which assumes the corona to be current-free.
This assumption leads to a well-posed mathematical problem
whose solution is relatively straightforward.  Nevertheless, 
several versions of the potential field are possible depending 
on the treatment of the
boundaries.  For example, the magnetogram of a single
active region can be extrapolated onto a finite computational grid,
into an infinite half space, or into a spherical corona inside a source
surface.  Each choice has advantages and all are in common use.  Since
these fields are all different, it is to be expected that each will
produce a different connectivity.

In this work we will explore the difference in connectivity produced
by different methods of field extrapolation.  Since the principles of 
potential field extrapolation are so simple and well understood, we 
restrict our investigation to these alone.  We will
explore the differences produced by different treatments of the
boundaries when making potential field extrapolations.  In addition to
their many other complications, sophisticated extrapolation methods,
such as the NLFFF, must choose between these same boundary conditions.
It is therefore worth quantifying the effect of these
choices on connectivity before considering the effects of more
complex extrapolations.

When a coronal field model is to be used only to compute connectivity
between unipolar photospheric regions, it is possible to replace those
regions by magnetic point charges.  The result, known as a 
magnetic charge topology model (MCT), is a kind of field commonly
used to study magnetic topology 
\citep{Baum1980,Gorbachev1989,Brown1999b} as well as
to quantify connectivity in observed fields
\citep{Longcope1998,Longcope2005,Longcope2007}.
Point magnetic charges
situated on the photospheric level create certain unphysical
artifacts, such as divergent magnetic field.  These artifacts are
absent from more traditional extrapolations which take the magnetogram
itself for the lower boundary condition.  MCT models do,
however, offer the advantage that their topologies may be
rigorously and systematically characterized
\citep{Longcope2002b,Beveridge2005}.  Furthermore, a potential
MCT field in an unbounded half space takes the form of an analytic
sum whose evaluation does not require a computational grid.

While it is clear that point charges introduce tremendous errors in
local properties of the magnetic field, it is not clear how much they
affect its global connectivity.  Away from a given unipolar region,
the potential field will be dominated by its lowest 
multipole moments, monopole and dipole.  These terms are exactly
matched by a single point charge.  In this work we quantify the
difference in connectivity produced by using point charges in a
potential field extrapolation.

In order to make a realistic comparison of connectivities we use a
magnetogram of an actual active region.  This magnetogram is analyzed
using the \citet{Barnes2005} partitioning algorithm with a range of
different parameters.  The resulting partitions consist of anywhere
from 35 to 395 different unipolar regions.  We compute connectivities
between these regions using four different potential magnetic field
extrapolations.  We find that the connectivity of a given partition 
varies by no more than 15\% regardless of what potential field
extrapolations is used.  Among the extrapolations, the choice of outer
boundary makes the greatest difference.  The use of point sources
changes the connectivity by roughly 6\% (in this case) regardless of
how many unipolar regions are present.

The next section reviews the process of partitioning and shows how its
parameters affect the result.  Section 3 defines the connectivity flux
and describes our method for calculating it.  The following section
describes the different extrapolation methods we explore.  The
connectivities are then compared in section 5.

\section{Partitioning the magnetogram}

\begin{figure}[htp]
\epsscale{1.0}
\plotone{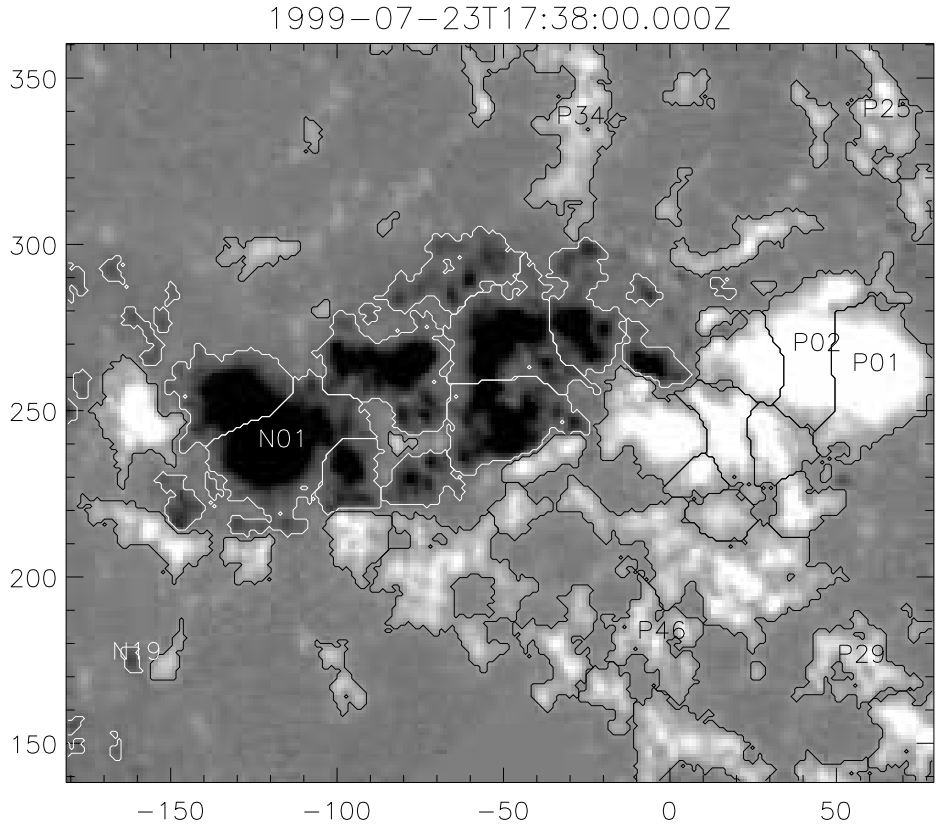}
\caption{The vertical field, $B_z(x,y)$ from
the IVM magnetogram of AR 8363 (grayscale).  Axes give
coordinates in arcseconds from disk center.  Curves outline
the regions from a partition with saddle level of $B_{\rm sad}=100$ G and
smoothing of $h=1.0$ pixel.  Several regions are labeled for future
reference.}
	\label{fig:msk}
\end{figure}

We begin with the single magnetogram
from the Imaging Vector Magnetograph \citep[IVM]{Mickey1996}
shown in \fig\ \ref{fig:msk}.  The
magnetogram is of AR 8636 from 23 July 1999, 
and includes most of the flux obviously
belonging to the active region (AR).  The inversion of the 
spectra to produce magnetic field maps is described in \cite{params}, 
while the ambiguity inherent in the observed transverse component 
of the field was resolved using the method described in \cite{precip1}. 
The three vector components of
the resulting vector magnetic field are used to 
compute the vertical (i.e.\ radial) component in each pixel, $B_z(x,y)$.
These values form a $237\times202$ 
array of $1.1''\times1.1''$ pixels within the plane of the sky.
Since the active region is relatively close to disk center, and we are
using the magnetogram for illustration  purposes, we do not project
the image onto the solar surface.  Instead we perform all analysis
within the plane of the sky.

\begin{figure}[ht]
\epsscale{0.7}
\plotone{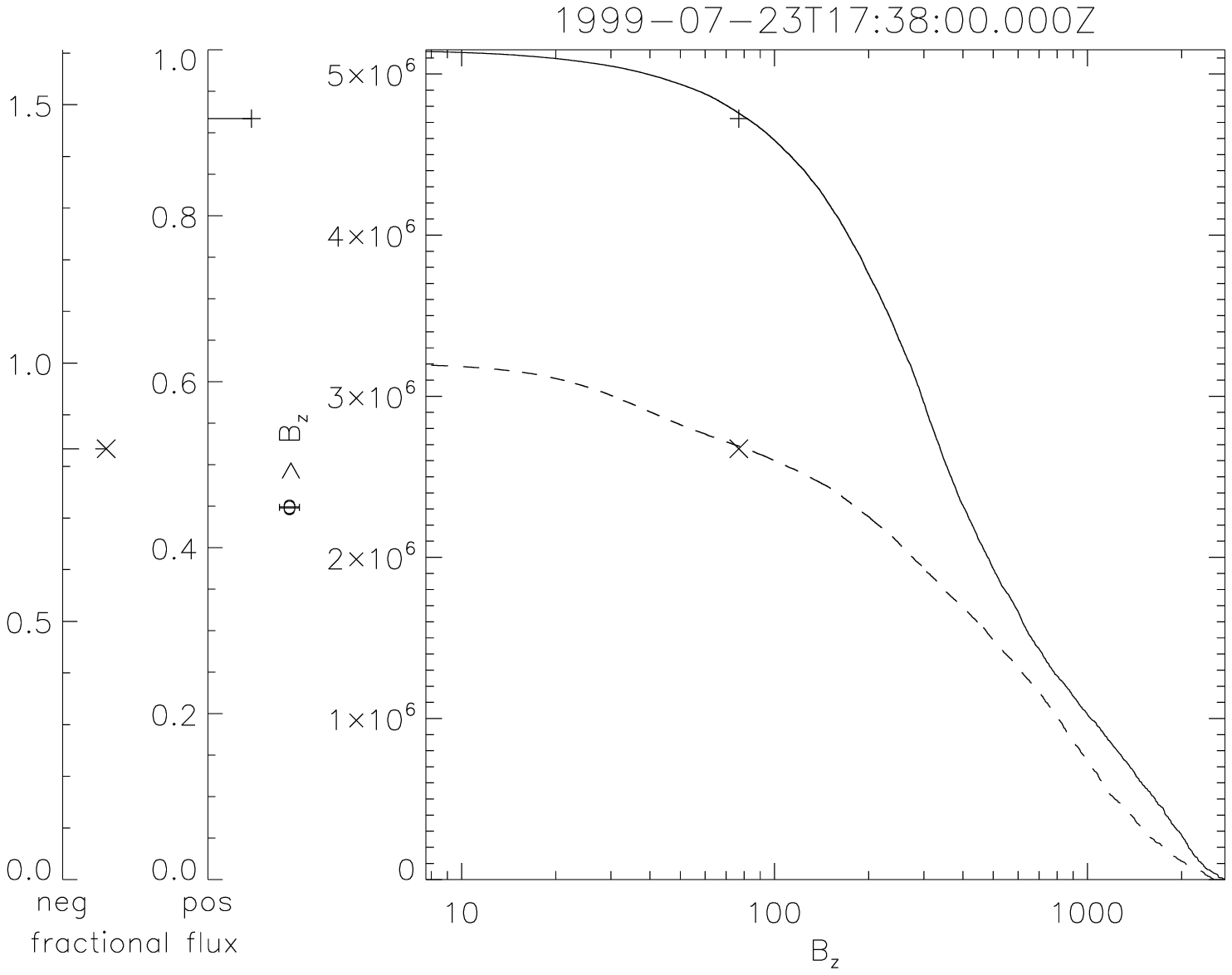}
\caption{A cumulative histogram of the IVM magnetogram from \fig\
\ref{fig:msk}.  Solid and dashed lines show the amount of magnetic
flux in the plane-of-the-sky (in G$\times$ square-arc-seconds) 
above a given field strength, in positive and
negative polarity respectively.  The plus and times mark the amount
included within regions; the threshold is $B_{\rm th}=76.9$ G.}
	\label{fig:msk_hist}
\end{figure}

It is evident from the cumulative histograms of positive and
negative pixels, shown in  \fig\ \ref{fig:msk_hist}, that the data are
dominated by positive 
flux.  Positive pixels ($B_z>0$) account for $\Phi_+=5.1\times10^6$ G
arcsec$^2$, while negative pixels compose less than two thirds of
these values $\Phi_-=3.2\times10^{6}$ G arcsec$^2$.  
(Had the radial field been mapped to
the solar surface the fluxes would have been $2.9\times10^{22}$ Mx and
$1.8\times10^{22}$ Mx respectively.)   It can be seen from the histogram
that field stronger than 500 G, which accounts for $\simeq2\times10^6$ G
arcsec$^2$, is much better balanced; most of the excess positive 
flux is weaker than this.  
This apparent imbalance
probably arises from the exclusion from the IVM field of view, of
an extended, diffuse region of negative polarity to the East.
This
is possibly part of an older, decaying AR into which 8636 emerged.
There is also an excluded region of more positive polarity to the
South of the IVM field of view.

Some degree of flux imbalance is inevitable in any magnetogram data.
Consequently, any method of magnetic extrapolation and any
determination of magnetic connectivity, must somehow accommodate
imbalance.  Our example, with its extreme degree of imbalance, will
bring these issues to the fore.  Moreover, we show below that
connections outside the AR are quantified more accurately and
with less computation
in cases of very strong imbalances.  It is for
these reasons that we select the IVM data from \fig\ \ref{fig:msk} for
the present study.

The magnetogram is next subjected to a process called 
{\em partitioning} \citep{Barnes2005,Longcope2007b} whereby pixels are
grouped into
unipolar regions.  Pixels with field strength below a cutoff, 
here set to the $3 \sigma$ noise level, $B_{\rm th}=76.9$ G, 
are discarded, and the remaining pixels are grouped using a
gradient-based tessellation scheme \citep{Hagenaar1997}.  This grouping
uses the gradient of a field constructed by convolving
$B_z$ with the kernel function
\be
  K_h(x,y) ~=~ {h/2\pi\over(x^2+y^2+h^2)^{3/2}} ~~.
	\label{eq:K}
\ee
This function integrates to unity over the entire plane, 
and to $\sqrt{1/2}$ within a circle of radius $h$.  It therefore
smoothes out fluctuations in $B_z$ on scales smaller than $\sim h$.
The function $K_h$ was chosen because it is the
Green's function for potential extrapolation upward from an unbounded
plane to a height $h$.   The convolution $K_h*B_z$ therefore 
resembles the vertical field within the plane $z=h$.

The gradient-based tessellation assigns a unique region label to every
local maximum in the smoothed field $|K_h*B_z|$ for which 
$|B_z|>B_{\rm th}$.  The label from a given maximum is given to all
pixels which are strictly downhill with 
respect to $|K_h*B_z|$, and also have $|B_z|>B_{\rm th}$.  
The resulting regions
are separated by areas where $|B_z|<B_{\rm th}$, or by internal
boundaries emanating from saddle points in the convolution
$|K_h*B_z|$.   The next step, called saddle-merging, eliminates any
internal boundary at whose saddle  point $|B_z|$ is greater than a
value ${\rm min}(|B_{\rm pk}|)-B_{\rm sad}$, where $B_{\rm pk}$ are the
values at the neighboring peaks, and $B_{\rm sad}$ is a parameter of
the partitioning.  The regions are merged by relabeling
the smaller one with the region number of the larger.  Of the
remaining regions, any which have flux less than $10^3$ G arcsec$^2$
are discarded.

\begin{figure}[htp]
\epsscale{1.0}
\plotone{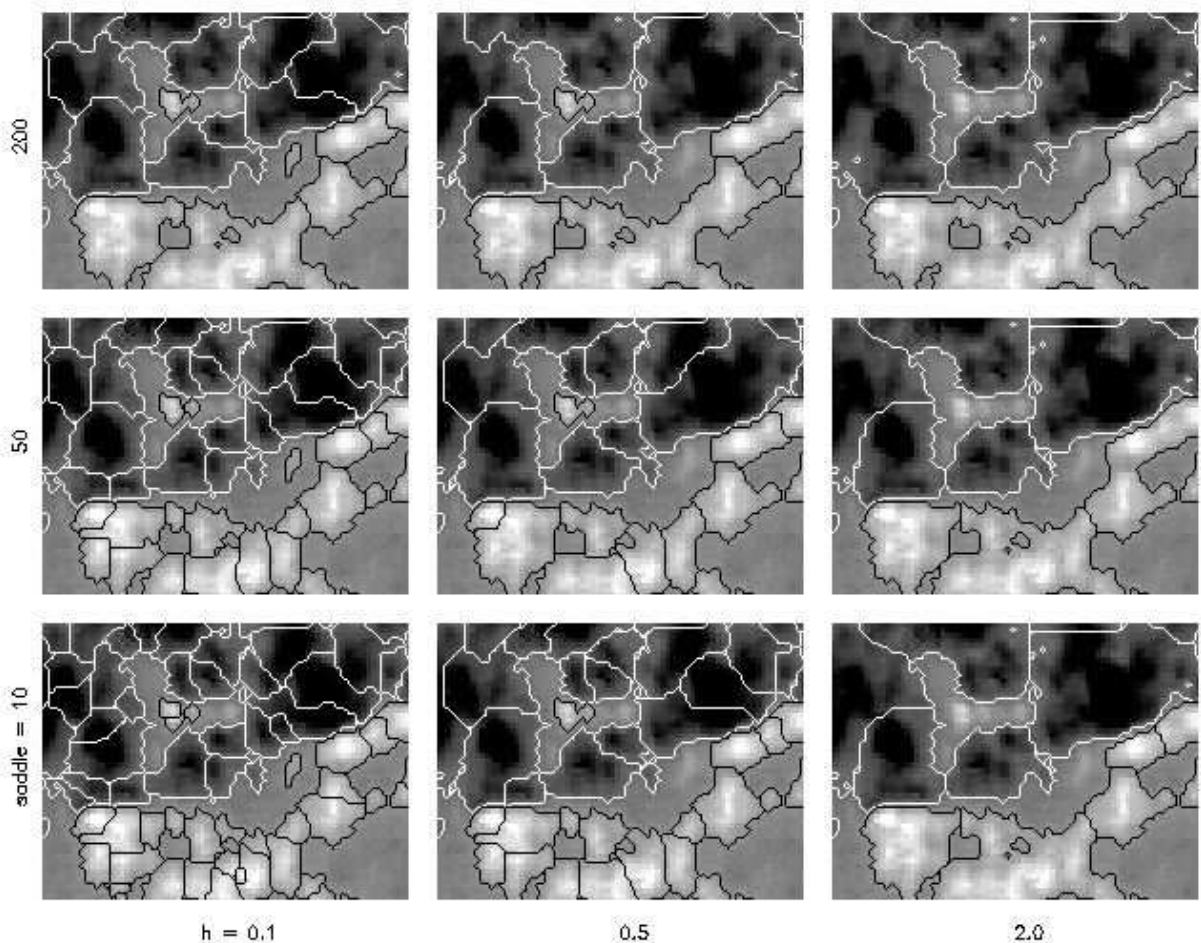}
\caption{A subsection of the magnetogram (see \fig\ \ref{fig:msk})
partitioned using different parameters.  The columns show the 
smoothing parameters $h=0.1$ pixel (left),
$0.5$ (center) and $2.0$ (right).  The rows show the
saddle-merging levels $B_{\rm sad}=10$ G (bottom), $50$ G (middle) and
$200$ G (top).  Note that increased smoothing removes the small positive 
polarity area surrounded by negative polarity, while increased saddle-merging 
simplifies it to a single partition without removing it. }
	\label{fig:part_sweep}
\end{figure}

The partition of a particular magnetogram depends critically on the
parameters $h$ and $B_{\rm sad}$, as illustrated by \fig\
\ref{fig:part_sweep}.  Increasing the smoothing-kernel width $h$
diminishes the number of local maxima.  The result is fewer
regions which are consequently larger.  Similarly, increasing $B_{\rm
sad}$, eliminates more internal boundaries, again yielding
fewer regions.  The progression is evident in \fig\
\ref{fig:part_sweep} by scanning up the columns or rightward along the
rows.  The parameters on the upper right
($h=2.0''$, $B_{\rm sad}=200$ G) 
partition the entire magnetogram into in $N=40$ 
regions, while those in
the lower left ($h=0.1''$, $B_{\rm sad}=10$ G) partition it into
$N=395$.

Region $\reg_a$ from a given partitioning is a set of pixels in which
$B_z$ is of the same sign: the region is unipolar.  The region may be
characterized by its signed net flux
\be
  \Phi_a ~=~ \int\limits_{\reg_a} B_z(x,y)\, dx\, dy ~~,
\ee
and its centroid location
\be
  \bar{\xvec}^a ~=~ \Phi_a^{-1}
  \int\limits_{\reg_a} \xvec\, B_z(x,y)\, dx\, dy ~~.
\ee
(We write integrals for mathematical clarity, but these are
actually computed as sums over pixels in $\reg_a$ multiplied by the
pixel area $A_{\rm pix}=1.21$ arcsec$^2$.)  

Further characterization of a region is provided by its quadrupole
moment
\be
  Q^a_{ij} ~=~ \Phi_a^{-1} \int\limits_{\reg_a}
  (x_i-\bar{x}^a_i)(x_j-\bar{x}^a_j)\, B_z(x,y)\, dx\, dy ~~,
\ee
where $i$ and $j$ are component indices for the horizontal vectors
(either $1$ or $2$).  One measure of a region's horizontal
extent is its radius of gyration
\be
  r_g^a ~=~ \sqrt{ Q^a_{11} + Q^a_{22} } ~~.
\ee
Its elongation may be characterized by
\be
  \varepsilon^a ~=~ 1 - {\lambda^a_{<}\over\lambda^a_{>}} ~~,
\ee
where $\lambda^a_{<}$and $\lambda^a_{>}$ are the smaller and larger
eigenvalue of $Q^a_{ij}$.  Since $Q_{ij}$ is positive definite the
elongation parameter will lie in the range $0\le\varepsilon<1$.  
An axi-symmetric flux distribution will have $\varepsilon=0$, while a
very long distribution will have $\varepsilon\simeq1$.

\begin{figure}[htp]
\epsscale{0.8}
\plotone{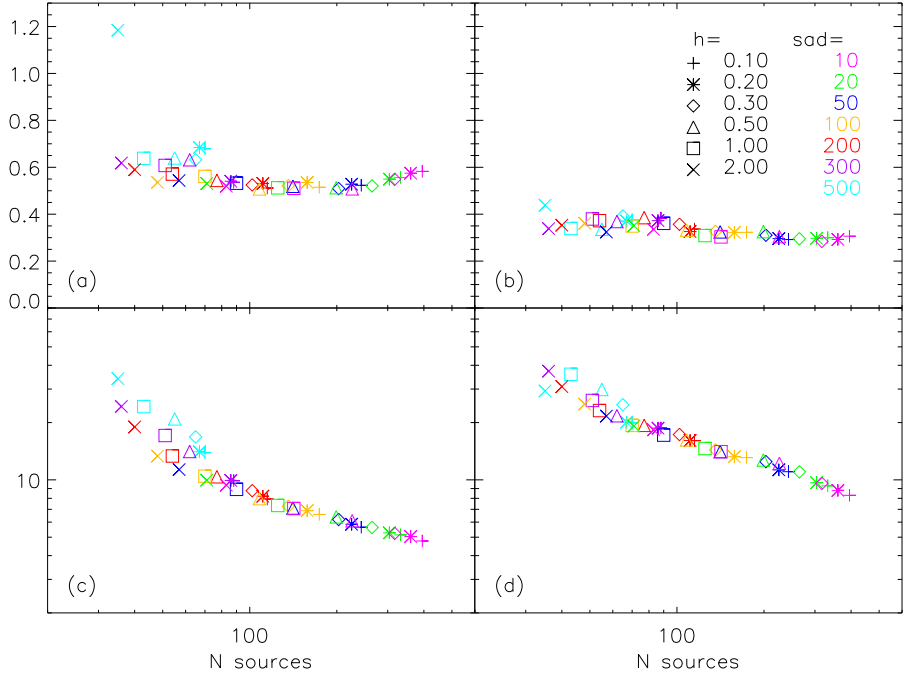}
\caption{Flux-weighted averages of region characteristics for
different partitions.  The bottom panels plot the radius of gyration 
$r_g$ (c) and the distance to nearest neighbor, $\Delta_n$ (d), both
expressed in arcseconds.  The top panels
show the ratio $r_g/\Delta_n$ (a) and the region elongation $\varepsilon$ (b).
In each plot the value is plotted against the
total number of regions in that partition.  The different symbols
denote different smoothing parameter and colors designate the
saddle-merging level.  The key for all four plots appear in the upper
right of (b).  An orange square ($B_s=100$ G and $h=1.0$)
marks the partition from \fig\ \ref{fig:msk} with its $N=70$ regions.}
	\label{fig:part_smry}
\end{figure}

For a given choice of parameters, $h$ and $B_{\rm sad}$, the partitioning
algorithm will
break a magnetogram into regions with various characteristics,
$\Phi_a$ and $r_g^a$.  The partitioning may be characterized as a
whole using a flux-weighted average of each characteristic
\be
  \avg{f} ~=~ {\sum\limits_a|\Phi_a|f_a\over\sum\limits_a|\Phi_a|}
   ~~.
	\label{eq:avg}
\ee
Figure \ref{fig:part_smry} shows the flux-weighted averages of 
quantities arising from partitions with different parameters.  Even as
the parameters cover a rectangle of $h$ and $B_{\rm sad}$
most of the averaged values fall on a single curve ordered by the
total number of regions $N$.  The average elongation,
$\avg{\varepsilon}$ (b) decreases slightly from $0.4$ to
$0.3$ as $N$ increases from $40$ to $400$.  It would seem the finer
partitionings (smaller $h$ or smaller $B_{\rm sad})$ produce less
elongated regions.

The average radius of gyration (c) shows a 
far more pronounced decrease with increasing number.  Clearly finer
partitioning produces regions which are generally smaller.  The
panel to its right, (d), plots the distance, denoted $\Delta_a$, from
the centroid of region $a$ to the nearest neighboring centroid of
either polarity.  Its flux-weighted average falls along the
curve $\avg{\Delta_a}=171N^{-1/2}$.  Had the centroids been scattered
randomly over the magnetogram their nearest-neighbor distance would
tend toward $\avg{\Delta_a}=120N^{-1/2}$, at large $N$ \citep{Kendall1963}.  
The median value of $\Delta_a$ does, in fact, approximate this curve, but the
flux-weighted mean is dominated by larger regions which tend to be
further from their neighbors.  This tendency leads to the larger
coefficient.

\section{Connectivity}

\begin{figure}[htp]
\epsscale{1.0}
\plotone{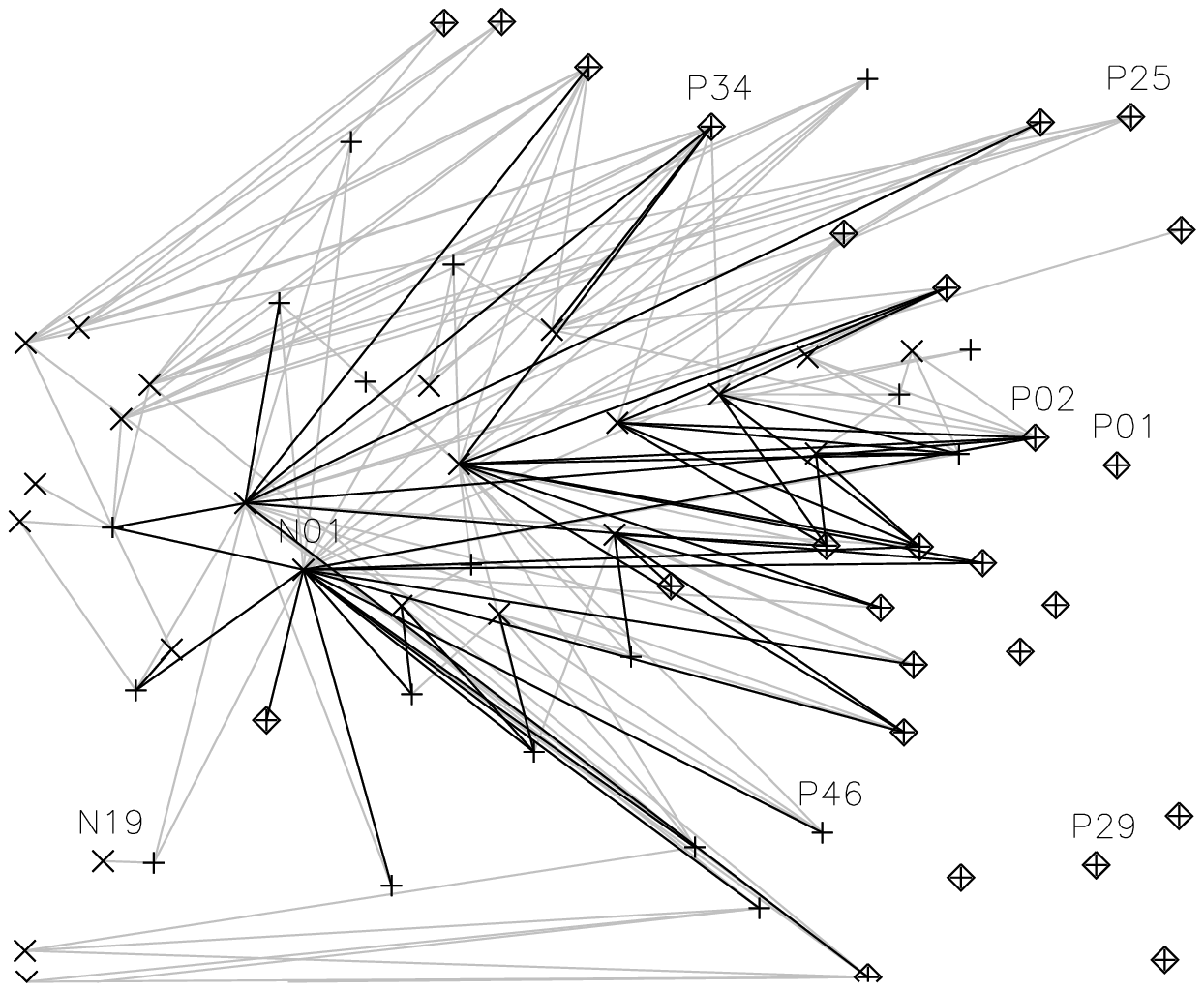}
\caption{A schematic diagram of the connections between source regions
from \fig\ \ref{fig:msk}, induced by coronal field $MB$.  
Centroids of positive and negative regions
are denoted by pluses and times.  Connections with flux in excess of
$0.25\%$ of $\Phi_{\rm tot}$ 
are represented by dark lines, those with less by
light lines.  Diamonds surround regions with connections to infinity.}
	\label{fig:cm_demo}
\end{figure}

The connectivity between regions can be found given some
coronal field anchored to the partitioned photospheric field.  There
is a connection between source regions $a$ and $b$ if a coronal field line
has one foot in positive region $\reg_a$ and the other in
negative region $\reg_b$.\footnote{We make a distinction here
between a connection and a {\em domain}, since it is possible for a pair
of sources to be connected through more than one domain 
\citep{Beveridge2005}.  Regardless of how many domains connect the
sources, we count this as a single connection.}  Figure
\ref{fig:cm_demo} is a schematic depiction of all connections produced
by a particular field anchored to the partitioning from \fig\
\ref{fig:msk}.

The connection between region $\reg_a$ and $\reg_b$ can be quantified by
the connection flux, $\psi_{ab}$
(the first index will always designate the
positive source region).  If a connection exists between these
regions then $\psi_{ab}>0$; if they are unconnected then
$\psi_{ab}=0$.  All of the flux anchored to negative region $\reg_b$
must have originated in some positive region, so
\be
  |\Phi_b| ~=~ \sum_{a\in \regs_+} \psi_{ab} ~~,
	\label{eq:Phib}
\ee
where $\regs_+$ is the set of all positive regions.  A similar
expression holds for positive region $a$
\be
  \Phi_a ~=~\sum_{b\in \regs_-} \psi_{ab} ~~,
	\label{eq:Phia}
\ee
except that $\regs_-$ includes an extra source to account
for field lines extending to ``infinity'' due to the net positive
imbalance.  This new source has flux
$|\Phi_{\infty}|=\Phi_+-|\Phi_-|$, in order to account for all flux which
cannot close in a photospheric negative source.
With the inclusion of infinity as a negative source, 
sources of each sign will have the same net flux
\be
  \Phi_{\rm tot} ~=~ \sum_{a\in \regs_+} \Phi_a 
  ~=~ -\sum_{b\in \regs_-} \Phi_b ~=~
  \sum_{a\in \regs_+}\sum_{b\in\regs_-}\psi_{ab} ~~.
	\label{eq:Phitot}
\ee
This is the total amount of flux in the field.

The connection flux $\psi_{ab}$ can be estimated using a Monte Carlo
method \citep{Barnes2005}.  A number $n_a$ of footpoints are selected
randomly within positive source region $\reg_a$.  
Each field line is then followed
to its other end (or to a distance from which
which it will not return).  The number which end at region $\reg_b$ is
designated $m_{ab}$.  Field lines are then randomly initialized from
negative regions (except infinity) and traced backward to their
positive source region.  The number initiated in $\reg_b$ which
``terminate'' in $\reg_a$ are denoted $m_{ba}$.  The Bayesian estimate
of the connection flux $\psi_{ab}$ \citep{Barnes2005} combines the
information from tracing in both directions as
\be
  \psi_{ab} ~=~ {m_{ab} + m_{ba} \over {n_a/\Phi_a + n_b/|\Phi_b|}}
  ~~.
	\label{eq:MCest}
\ee
Since no points were initiated at infinity, $n_b=0$ and
$m_{ba}=0$ for that case ($b=\infty$).  The estimate then reduces to
$\psi_{ab}=(m_{ab}/n_a)\Phi_a$.

Expression (\ref{eq:MCest}) is a Monte Carlo estimate of the actual
connection flux so it will include some statistical error.  This error can be
estimated using the (approximately) Poisson statistics of the counts
$m_{ab}$ and $m_{ba}$.  If an actual connection includes very little
flux it is possible that none of the randomly generated
field lines will sample it and
$m_{ab}=m_{ba}=0$.  The probability that an actual connection will be
erroneously missed for this reason is
\be
  P_{\rm miss}(\psi_{ab}) ~=~ e^{-\psi_{ab}/\tilde{\phi}_0}  ~~,
\ee
where $\tilde{\phi}_0=1/(n_a/\Phi_a + n_b/|\Phi_b|)$.  
Naturally the use of more
points (i.e.\ larger values of $n_a$ and $n_b$) will make this
increasingly unlikely.  Nevertheless, there will always be a
possibility that some number of very small connections have remained
undetected.  For one particular case (later defined as field $PH$) a
topological analysis of the connections revealed that 10 of 109
connections were erroneously missed by the Monte Carlo method.

% It is
% also noteworthy that the estimates (\ref{eq:MCest}) will not, in
% general, satisfy conditions (\ref{eq:Phib}) or (\ref{eq:Phia}).  Their
% level of violation provides an estimate of the error in the
% estimates of connection fluxes.

\section{Coronal fields}

\subsection{Different potential fields}

In order to define connectivity it is necessary to compute the entire
coronal field anchored to the photospheric field $B_z$.  Although we
have chosen to
restrict consideration to potential fields,
\hbox{$\nabla\times\bvec=0$}, there are several different ways to 
compute a potential field from magnetogram data.   We consider a variety of
these fields and study the effect on the connectivity of the
differing fields.

One option is to compute the potential field within a
rectangular $L_x\times L_y\times L_z$, 
box, $\bvol$, with a lower boundary at the magnetogram,
$z=0$.  The four lateral boundaries are perfect conductors
($\nhat\cdot\bvec=0$) positioned along 
the edges of the magnetogram.  The vertical field at the lower
boundary, $B_z(x,y,0)$, is taken from the magnetogram 
and is therefore not balanced.   
An equal net flux
must cross the upper boundary or no solution would be possible for
which $\nabla\cdot\bvec=0$.
We achieve this with a uniform vertical field
along the upper boundary: $B_z(x,y,L_z)=|\Phi_{\infty}|/(L_xL_y)$.  A
field line crossing this upper boundary is designated as a
connection to infinity.  We choose to place this upper boundary at
$L_z=220''$, approximately equal to $L_y$, and slightly less than
$L_x$.

The alternative to the computational box $\bvol$ 
is to use a coronal field extending
throughout the entire half-space $z>0$.  Such an unbounded
field is computed, in principle, by
convolving the field at $z=0$ with a Green's function for a point
magnetic charge at $z=0$.  In practice we compute either a portion of
the field on a grid or compute it along
a field line as we trace it.
We distinguish between the box boundary and the half-space
using superscripts $B$ and $H$ respectively.

For the purposes of computing connections between
unipolar regions it is possible to replace each region with a point
source.  Region $\reg_a$ is replaced with a magnetic charge of
strength $q_a=\Phi_a/2\pi$ at position $\bar{\xvec}^a$ on the
photospheric plane, $z=0$.  This matches a
multipole approximation of the potential field from $\reg_a$
up to the dipole term, thus it is expected
to be accurate at distances $|\xvec-\bar{\xvec}^a|\gg r_g^a$.  
Fields computed using point sources will be assigned a superscript
$P$.  With this simplification the convolution required for the
half-space computation becomes
\be
  \bvec^{\rm (PH)}(\xvec) ~=~\sum_a {\Phi_a\over2\pi}\,
  {\xvec-\bar{\xvec}^a\over|\xvec-\bar{\xvec}^a|^3} ~~,
	\label{eq:BPH}
\ee
where the sum is over all $N$ sources, not including infinity.

An alternative to photospheric point sources is
to compute a potential field
matching the magnetogram pixel-for-pixel.
Field vectors are computed on
a three-dimensional uniform cartesian 
grid, for example within $\bvol$.  In order that each field line have a
defined connectivity magnetogram pixels which do not belong
to a source region are set to zero.  The resulting magnetograms differ
slightly for different partitionings, but for most cases
$\Phi_+=4.7\times10^6$ G arcsec$^2$ and
$\Phi_-=2.7\times10^6$ G arcsec$^2$.  Fields anchored in this way are
designated by a superscript $M$.

The field $\bvec^{(MB)}$, bounded by conducting boundaries, is readily
computed on the Cartesian grid using Fourier methods.  The
extrapolation into the half-space, $\bvec^{(MH)}$, is done onto
a larger Cartesian grid which we call
$\mhvol$.  The field at any grid point can be found by convolving the
entire magnetogram with the half-space Green's function.  Performing
the convolution for every grid-point is very time-consuming so we use
it only for points along the boundary $\partial\mhvol$.  We then
use the efficient Fourier techniques to compute the interior potential field
matching these boundary values.  It is noteworthy that flux crosses
the boundaries $\partial\mhvol$, so these 
are not genuine boundaries of the field.

\subsection{Calculating the connection fluxes}

For each type of coronal field, connection fluxes $\psi_{ab}$ are
computed using the Monte Carlo estimate in eq.\ (\ref{eq:MCest}).
Approximately $n_a=|\Phi_a|/(2\phi_0)$ field lines are initiated on
source $a$, where we have chosen $\phi_0=10\,$ G arcsec$^2$.  The
smallest flux reported will be $\psi_{ab}\simeq\phi_0$, for example
when $m_{ab}=1$ and $m_{ba}=0$.  With our choice of $\phi_0$, every
source will have at least 50 lines, and the largest will have
$n_{P01}\simeq4\times 
10^4$.  One complete estimate, such as the one shown in \fig\
\ref{fig:cm_demo} requires 
$(\Phi_++|\Phi_-|)/20\simeq 3.7\times10^5$ field lines be
traced.

Field line initialization is different for the cases anchored to
point sources than for those anchored to magnetograms.
In the point-source cases, the $n_a$ points for a given source
are randomly generated with a
uniform distribution over a very small hemisphere centered at the
point source.  The radius of the hemisphere is set to be 
small enough that the magnetic field is directed roughly 
radially outward (inward) from the positive (negative)
source.

For magnetogram cases the extended region $\reg_a$
is a set of $P_a$ pixels.
First, a list of $n_a$ pixels are randomly generated
so as to sample each pixel in $\reg_a$ with probability proportional
to its field strength ($p=A_{\rm pix}B_z/\Phi_a$).  
Since $n_a\gg P_a$, typically, 
the list will include many duplicate pixels.
For each random pixel (including all duplications) an initial point is
randomly generated with a uniform distribution over the pixel.  Thus no
two field lines from the same pixel will begin at the same point.  The
subsequent field line integration uses tri-linear interpolation to calculate
$\bvec(\xvec)$, so different points within the same pixel will belong
to different field lines.

\begin{figure}[htp]
\epsscale{1.0}
\plotone{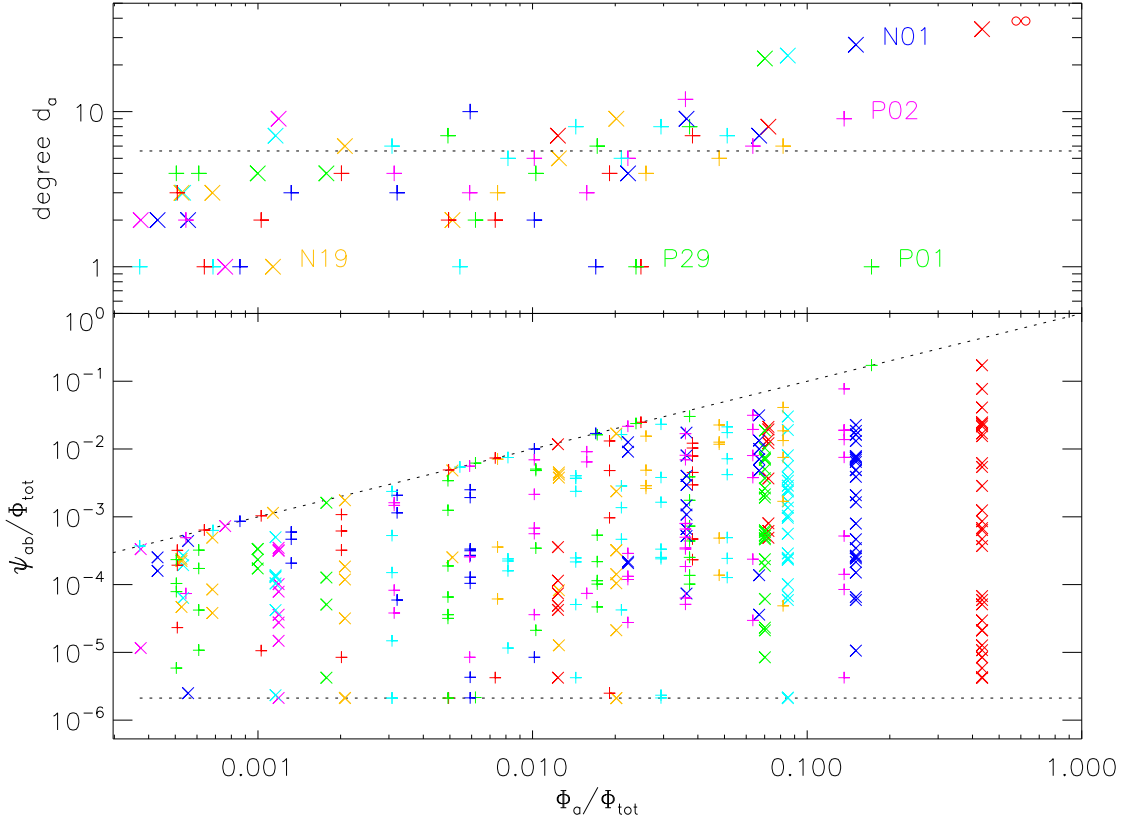}
\caption{The connection fluxes $\psi_{ab}$ from the field depicted in
\fig\ \ref{fig:cm_demo}.  The bottom panel plots the flux $\psi_{ab}$
against the flux of its sources, $\Phi_a$ (as a $+$) and
$|\Phi_b|$ (as a $\times$).  The sloped and horizontal
dotted lines show the maximum and minimum possible values:
$\psi_{ab}=\Phi_a$ and $\phi_0$ respectively.  
Colors are used to differentiate the
different sources, whose connections fall along a vertical line.
On each axis the fluxes are normalized
to the total, $\Phi_{\rm tot}$.  The top panel
plots the number of connections to each source, $d_a$.
Several of the sources are labeled for reference.  The
horizontal dotted line shows the mean degree, $\bar{d}_a=5.65$.}
	\label{fig:phi_cm_scat}
\end{figure}

Using these methods we estimate the connectivity of a given field
according to \eq\ (\ref{eq:MCest}).
As an illustration, consider the field $\bvec^{(MB)}$ 
from the partition shown in \fig\ \ref{fig:msk}.  Our estimate, shown
in \fig\ \ref{fig:cm_demo}, includes 198 different connections between its
$N=70$ sources (71 including $\infty$).
These connections are quantified by their
connection fluxes, $\psi_{ab}$, plotted in \fig\
\ref{fig:phi_cm_scat}.  The connections to a given source fall along a
vertical line below the diagonal, $\psi_{ab}=\Phi_a$ (dotted) in the
lower panel.  The number of connections to that source, called its
{\em degree} $d_a$, is plotted above its flux in the upper panel.

Some sources, such as $P01$ or $N19$, have only a single connection
($d_a=1$) and are called {\em leaves}.  In \fig\ \ref{fig:cm_demo},
leaves appear at the end of a single line ($N19$) or as an isolated
diamond ($P01$).  In the lower panel of \fig\ \ref{fig:phi_cm_scat},
the fluxes of a leaf connection naturally fall on the dotted diagonal
since all the flux from that source belongs to that single
connection.

In contrast to the leaf connections there are several sources, such as
$N01$ with many connections ($d_{N01}=27$).  Given $C=198$ connections
to $N+1=71$ sources, the average source must connect to
$\bar{d}_a=2C/(N+1)=5.65$ sources: the value marked by the 
dotted line in the upper panel.  (The factor of two arises from the
fact that each connection is incident on two different sources: one
positive and one negative).  There is a
notable tendency for larger sources, especially larger
negative sources, to have more connections.  As a consequence of this
tendency, the flux-weighted average, \eq\ (\ref{eq:avg}), of the
degree is $\avg{d_a}=9.35$ in this case.

\subsection{Connections to infinity}

In each different magnetic field there is open flux, represented by
field lines connected to infinity (formally a negative source).  In the
$MB$ and $PB$ fields, an open field line is one that terminates at the
upper surface of the box, $z=L_z$.  The 34 positive sources enclosed by
diamonds in \fig\ \ref{fig:cm_demo} are connected to infinity in the
$MB$ field for that partition.  These form the connections whose
fluxes fall in the right-most vertical row of $\times$s in \fig\
\ref{fig:phi_cm_scat}.

The $PH$ field occupies the entire half space and open flux
truly extends outward indefinitely.  Far from the AR the field
resembles that from a single point charge $|\Phi_{\infty}|/2\pi$; field
lines go outward approximately radially. 
There is a single separatrix surface dividing closed from open flux,
and once a field line has been integrated 
far enough to establish that it lies outside this
surface, it may be designated as a connection to infinity.

\begin{figure}[htp]
\epsscale{0.8}
\plotone{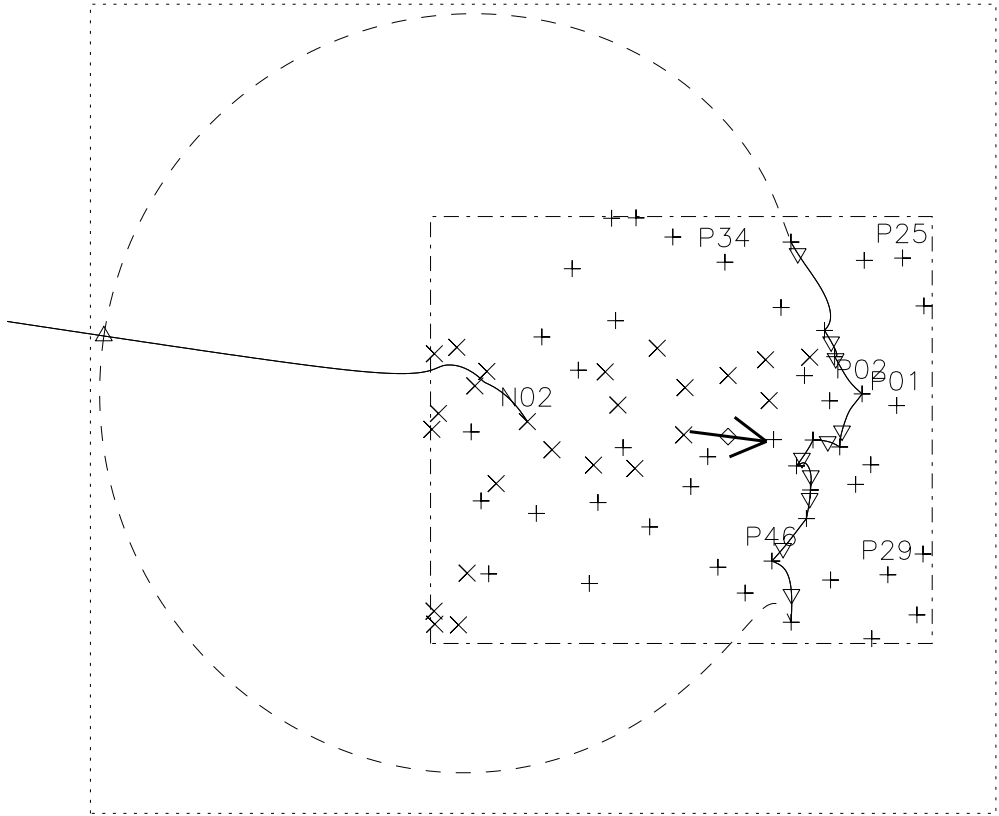}
\caption{The footprint of the separatrix dividing closed from open
flux (dashed and solid curves) in the field $\bvec^{(PH)}$ from the
partition of \fig\ \ref{fig:msk}.  Positive and negative sources are
denoted by $+$s and $\times$s respectively.  Triangles are magnetic
null points.  The triangle at the far left, lying along the dashed
curve, is the negative null whose fan surface is the separatrix.
The center of unsigned flux is denoted by a diamond, and the dipole
moment about this point is indicated by an arrow.  The broken
rectangle shows the extent of the magnetogram, and the larger dotted
one is the bottom surface of $\mhvol$.}
	\label{fig:dome}
\end{figure}

The separatrix between open and closed field is a dome 
anchored to a null point located $\sim 327$ arcsec from the center of
the AR (the triangle in \fig\ \ref{fig:dome}).  The footprint of the
dome passes along a series of spines (solid curves) linking positive
sources.  These sources connect to infinity as well as at least
one other negative source, $N02$, inside the dome.  Positive sources
outside the dome link only to infinity.

The dipole moment, computed about the center of unsigned flux (diamond
in \fig\ \ref{fig:dome}), is $\mu=5.1\times10^7$ G arcsec$^3$,
directed $7^{\circ}$ below ${\bf \hat{x}}$ as shown in the figure.  A
far field with this dipole and the net charge, 
$q_{\infty}=+|\Phi_{\infty}|/2\pi$ will vanish at one point located a distance
\be
  r_0 ~=~ 2{\mu\over q_{\infty}} ~\simeq~ 318\,{\rm arcsec} ~~,
	\label{eq:r0}
\ee
from the center of unsigned charge.  For a region with positive net
flux, as we have, the null is situated in the direction opposite to
the dipole moment.  Clearly this is the approximate location of the
null point whose separatrix divides open from closed flux in the $PH$
field.  Had the region been more balanced, $q_{\infty}$ would have
been smaller and the separation of open from closed flux would have
occurred much farther out.

There are 22 positive sources linked to infinity by dint of lying 
on or outside the separatrix in field $PH$.  In contrast, the $MB$,
field has 34 sources connected to $\infty$, even though there is the
same amount of open flux, $|\Phi_{\infty}|$, in each case.
Comparing the labeled sources in \figs\
\ref{fig:cm_demo} and \ref{fig:dome}, shows that different sources are
so connected in the $MB$ and the $PH$ fields.  These connections are
just some of the differences between the two fields, explored
further in the next section.

Establishing and quantifying connections to infinity
is particularly challenging for
the $MH$ field.  Although this field formally extends throughout the
half-space, it is known only on a Cartesian grid covering 
$\mhvol$.  A field line can therefore 
be followed to the boundary of $\mhvol$
but no farther.  It is, in principle, possible for field at
$\partial\mhvol$ to be directed both inward and outward.  When this
is the case there must be some field lines which leave the volume
where $B_n>0$ and return where $B_n<0$.  Naturally these field lines
cannot be followed, so their flux cannot be correctly assigned to a
connection.  Due to the large flux imbalance in our magnetogram we
were able to select a volume $\mhvol$ for which $B_n\ge0$ on 
all outer surfaces (see \fig\ \ref{fig:dome}).
Thus any field line encountering the boundary
is necessarily connected to infinity.  

In order to assure $B_n>0$ on the outer boundary 
it is necessary (but not sufficient) that
$\mhvol$ enclose the separatrix dome in the field $\bvec^{(MH)}$.  
This surface will closely resemble (but not exactly match) the
separatrix dome from $\bvec^{(PH)}$ since both fields approach the same 
far-field form.  It is evident from \fig\ \ref{fig:dome} that the base
of $\mhvol$ does enclose the latter separatrix.  This requirement means
$\mhvol$ must be considerably larger than $\bvol$, so computations in
$\bvec^{(MH)}$ are much more expensive than for $\bvec^{(MB)}$ or
$\bvec^{(PB)}$.  If the active region had had better flux balance then
$\mhvol$ would need to be still larger and computation would have
become prohibitively expensive.

\section{Comparisons}

\subsection{Different fields from a single partition}

\begin{figure}[htp]
\epsscale{0.9}
\plotone{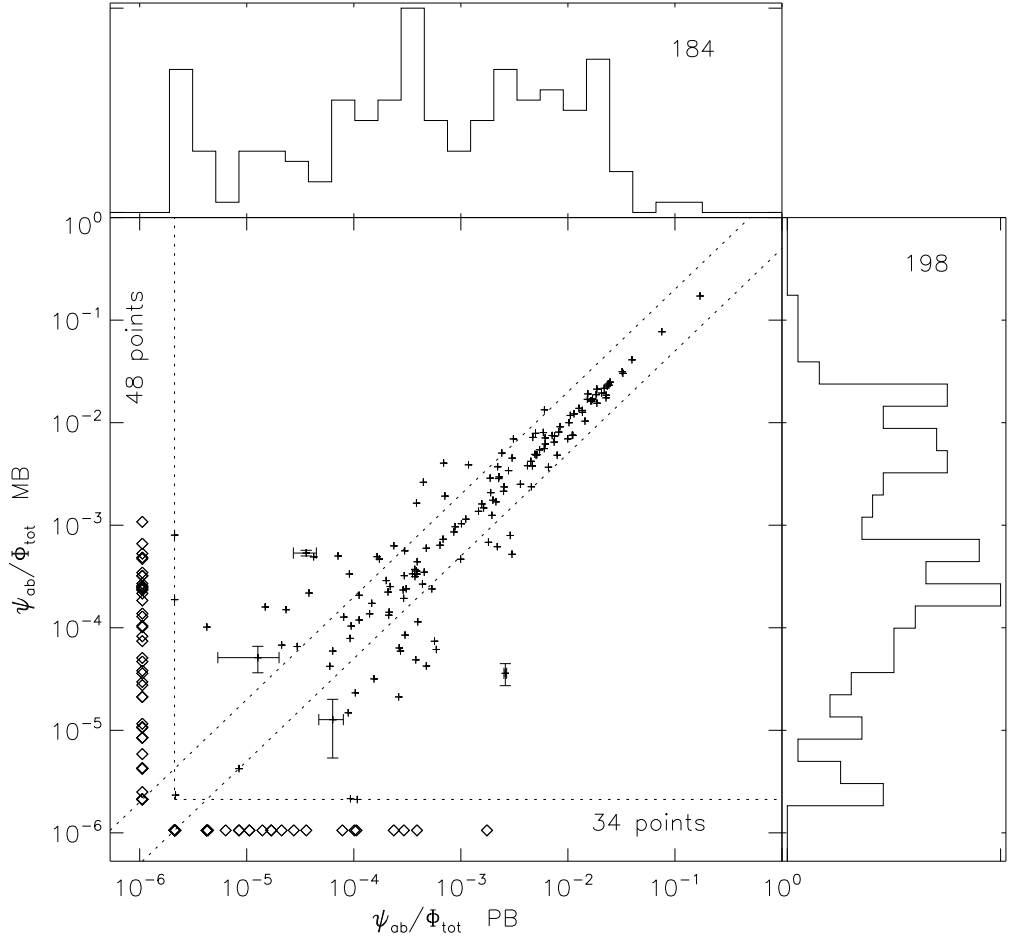}
\caption{The fluxes of identical connections from two different fields,
$MB$ (vertical axis) and $PB$ (horizontal axis) plotted against one
another.  Fluxes are normalized to the total $\Phi_{\rm tot}$. 
Vertical and horizontal dotted 
lines show the minimum flux, $\phi_0$.  Singlet connections (absent from
one of the fields) appear as diamonds below or to the left of these
limits.  For example, the vertical line of diamonds are absent from $PB$
but present in $MB$ with fluxes indicated by their position.  The
diagonal dotted lines mark $\psi_{ab}^{(MB)}=2\psi_{ab}^{(PB)}$
(upper) and $\psi_{ab}^{(MB)}=\psi_{ab}^{(PB)}/2$ (lower).
Statistical errors are indicated on a few representative points.
Plotted
along the right and top are histograms of $\ln\psi_{ab}$ for that
field. The number appearing in the histogram gives the total number
of connections.}
	\label{fig:cm_scat}
\end{figure}

Four different coronal fields can be generated from a single
partition in the fashion described above.
Each field will contain the same total flux,
$\Phi_{\rm tot}$, interconnecting the same $N+1$ sources.  The
connections will not, however, be the same for the different fields.
Figure \ref{fig:cm_scat} shows the connection fluxes $\psi^{(MB)}_{ab}$
(the ones from \fig\ \ref{fig:phi_cm_scat}) plotted against those
induced by $\bvec^{(PB)}$.  There are $C^{(MB)}=198$ connections in
the former and only $C^{(PB)}=184$ in the latter.  Moreover, there are
48 connections in $MB$ which do not appear in $PB$; these appear as
diamonds along a vertical line in the central panel.  Similarly 34
connections in $PB$ not present in $MB$ form the 
horizontal row of diamonds.
We refer to either of these as {\em singlet connections}.  The
remaining 150 connections, common to both fields, are plotted as $+$s
in the central panel.

The tendency for common connections ($+$s) to cluster about the diagonal,
especially at the upper right, shows that connections have similar
fluxes in both fields.  The connections falling outside the dotted
diagonals (i.e.\ disagreeing by more than a factor of two) are
overwhelmingly dominated 
by smaller connections: $\psi_{ab}<10^{-3}\Phi_{\rm tot}$.
These small connections also compose almost all of the singlet
connections in either field.  Indeed, a great many of the singlets are
so small ($\psi_{ab}\simeq\phi_0$) that they had a significant
probability of going undetected even in the field where they were
found.  These tiny connections account for most of the spikes at the
small-flux end of each histogram. 

The statistical errors from the Monte Carlo calculations are
relatively large for small fluxes (found by only a few field lines).
On the logarithmic plot, like \fig\ \ref{fig:cm_scat}, the error bars
are largest at the bottom or left.  Above a value of $\psi_{ab} \simeq
10^{-3}\Phi_{\rm tot}$ statistical errors are less that $0.1\%$ and
errors bars are smaller than the symbols.

The impression given by a comparison such as \fig\ \ref{fig:cm_scat}
is that in spite of their differences the two coronal fields induce
connections which are largely in agreement.  The differences appear
mostly in the very small connections.  While these are small, most of
them lie well above the detection limit, $\phi_0$, and thus represent
genuine differences.  The fact that the differences are in small
connections suggests that they will not be of great importance to a
model of the field.

\begin{figure}[htp]
\epsscale{0.8}
\plotone{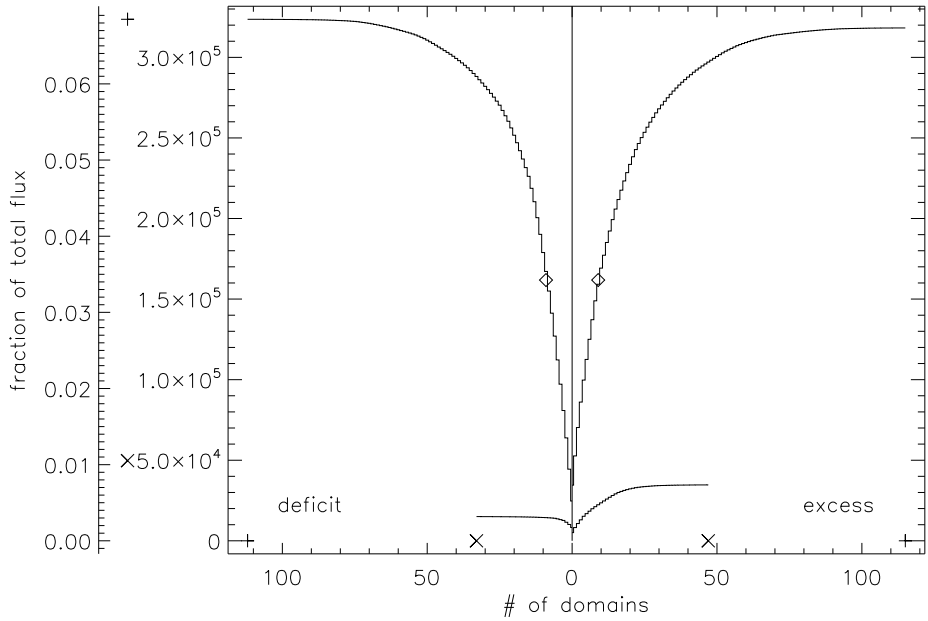}
\caption{Cumulative histograms of the flux differences
$\Delta\psi_{ab}$.  Those with $\Delta\psi_{ab}>0$ are graphed on the
right and the ones with $\Delta\psi_{ab}<0$ are on the left.  The
differences are sorted by decreasing magnitude, so the
upper curves plot the total difference for the $n$ largest differences of
that sign.  The right axis gives this flux in units of G
arcsec$^2$ while the left axis gives it in units of $\Phi_{\rm tot}$.
The lower curves are for only the singlet connections.  A $+$ and
$\times$ appears at the extreme values of the total and singlet
curves respectively.  These extreme values are also projected
to the ordinate and abscissa with the same symbol.  The sole
exception is that the $\times$ on the left gives the {\em sum} of the
total singlet fluxes of both signs, while the curves show the sums
for each sign separately. A $\diamond$ shows half the total discrepancy 
for each sign of $\Delta\psi_{ab}$.}
	\label{fig:fount1}
\end{figure}

In order to weight the most significant flux differences 
we focus on the difference
\be
  \Delta\psi_{ab} = \psi_{ab}^{\rm (MB)} - \psi_{ab}^{\rm (PB)} ~~.
	\label{eq:Deltapsi}
\ee
Connections with positive difference ($\Delta\psi_{ab}>0$) are those
for which  $MB$ has excess flux relative to $PB$; these
appear above the diagonal in \fig\
\ref{fig:cm_scat}.  Using \eq\ (\ref{eq:Phitot}) we can show that
\be
  \sum_{a\in \regs_+}\sum_{b\in\regs_-}\Delta\psi_{ab} ~=~ 0 ~~,
\ee
so there will be as much flux in connections with excess
($\Delta\psi_{ab}>0$) 
as in connections with deficit ($\Delta\psi_{ab}<0$).  Figure
\ref{fig:fount1} shows 
cumulative histograms of the flux differences of each sign.  In each
case the total discrepancy is $6.80\%$ of $\Phi_{\rm tot}$.
The connections are sorted by decreasing magnitude, so the histograms
rise sharply at first.  Ten to twelve connections account for half the
total discrepancy of each sign (shown by diamonds).  The rest of the
discrepancy occurs in the hundred or so other connections.

For a singlet connection, one of the terms on the right of \eq\
(\ref{eq:Deltapsi}) will vanish.  The magnitude of the difference will
therefore equal the other term.  The lower curves in \fig\
\ref{fig:fount1} show histograms formed from the singlet connections
alone.  The right curve accumulates the differences
in the 48 singlet connections in $MB$, while the left accumulates
the 34 singlets in $PB$ (for which $\Delta \psi_{ab}<0$).
Combining their totals yields $1.05\%\Phi_{\rm tot}$, so singlets
contribute only a small fraction to the overall flux discrepancy
($6.80\%$).

\begin{figure}[htp]
\epsscale{0.8}
\plotone{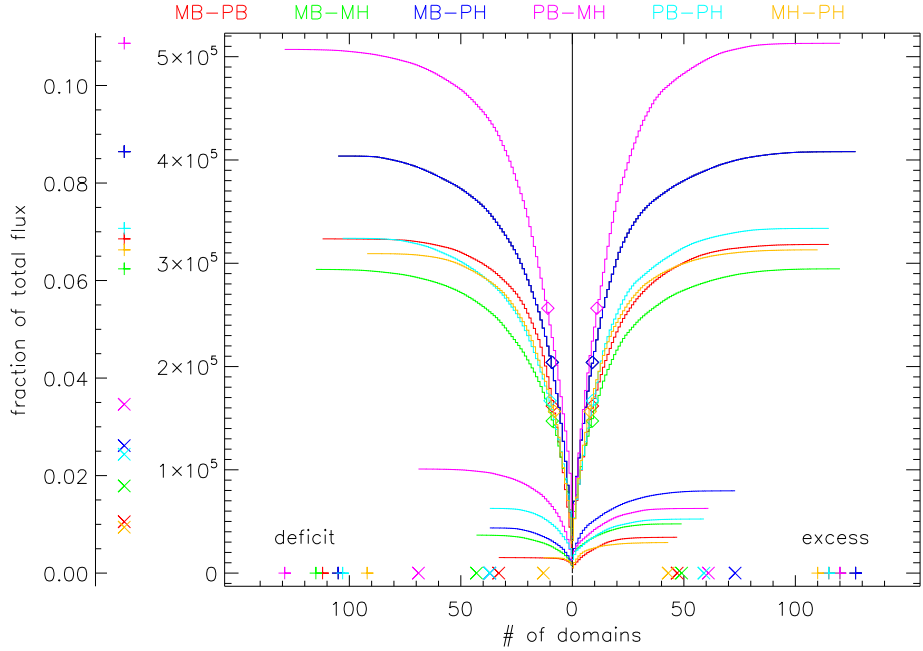}
\caption{A superposition of cumulative histograms comparing all four
different kinds of fields.  Each comparison is
the same as \fig\ \ref{fig:fount1}, but shown in a different color.
The color code appears along the top.}
	\label{fig:fount2}
\end{figure}

The inclusion of two other
kinds of coronal field leads to six different
pairwise comparisons of the kind just used.  
Figure \ref{fig:fount2} shows cumulative histograms like
those in \fig\ \ref{fig:fount1}, for all six possible flux differences.
All histograms have the same $\Upsilon$ shape as the ones in \fig\
\ref{fig:fount1}; a few connections account for the majority of the
discrepancy.  Variation among the three fields seems to be mostly a
matter of degree.

The largest discrepancies are between
fields differing both in the their anchoring (magnetograms
versus points) and their boundaries (box versus half-space): for
example between $PB$ and $MH$.  The fields differing in only one
respect, either anchoring or boundary, have histograms in a cluster
below the other two.  This tendency is true for both the total
histograms (upper curves) and singlet histograms (lower curves).

Making the same plots for many other partitions we find no exceptions 
to this $\Upsilon$ shape.  The histograms therefore
differ mostly in their magnitude, which may be summarized by the
maximum (marked by a $+$ along the left axes in \figs\
\ref{fig:fount1} and \ref{fig:fount2}).  We denote this value
\be
  \Delta\Psi_{X-Y} ~=~ 
  \half\sum_{a\in \regs_+}\sum_{b\in\regs_-}|\psi_{ab}^{(X)}-
  \psi_{ab}^{(Y)}| ~=~ \half\sum_{a\in \regs_+}\sum_{b\in\regs_-}
  |\Delta\psi_{ab}| ~~,
	\label{eq:DeltaPsi}
\ee
for a comparison between
$\bvec^{(X)}$ and $\bvec^{(Y)}$ for the same partition.  (Here $X$ and
$Y$ stand for any of $MB$, $PB$, $MH$ or $PH$.)  Thus \fig\ \ref{fig:fount2}
is summarized by the values $\Delta\Psi_{PB-MH}=10.8\%$,
$\Delta\Psi_{MB-PB}=8.6\%$, $\Delta\Psi_{MB-PB}=6.8\%$, and so forth (in
units of $\Phi_{\rm tot}$).

All the connection fluxes are calculated by Monte Carlo methods, so
the values of $\psi_{ab}^{(X)}$ and $\psi_{ab}^{(Y)}$ include
statistical errors.  The sum in (\ref{eq:DeltaPsi}) will therefore be
biased upward, and even an estimate of $\Delta\Psi_{X-X}$ will be
positive provided it uses
two different estimates of the connectivities from $\bvec^{(X)}$.  
(To see this note that
the sum in \eq\ [\ref{eq:DeltaPsi}] is over numbers which are
never negative and are usually positive due to errors.)
We can subtract the
expected bias, assuming errors in $\Delta\psi_{ab}$ to have Gaussian
distributions, following a procedure described in an appendix.   
Doing so for $\Delta\Psi_{MB-MB}$, for example,
yields a number consistent with zero.  Doing so for the
red curve in \fig\ \ref{fig:fount2} yields
$\Delta\Psi_{MB-PB}=6.6\%\pm 0.25\%$ (the value on the curve,
$\Delta\Psi_{MB-PB}=6.8\%$, is therefore biased upward by
$\simeq0.2\%$).  Thus there is a true difference between field $MB$
and $PB$, beyond that caused by statistical errors.

\begin{figure}[htp]
\plotone{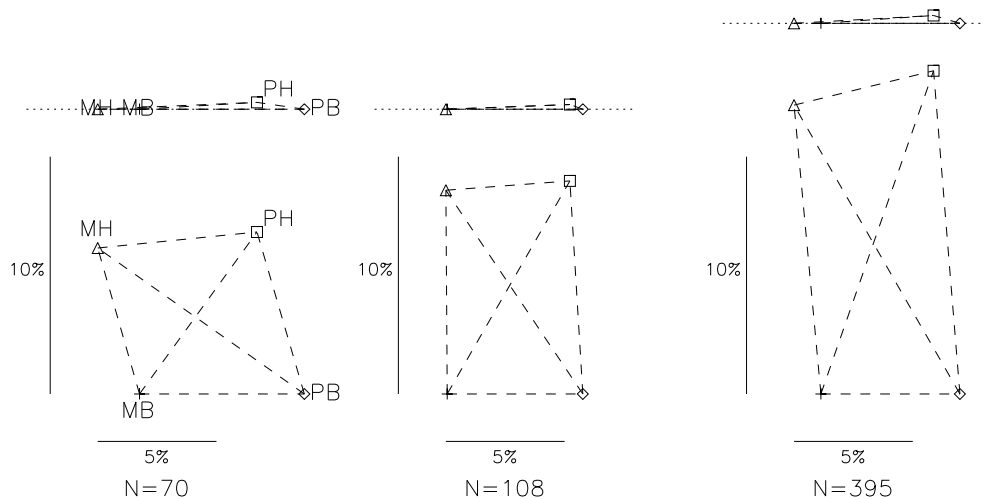}
\caption{Tetrahedra formed from the connectivity distances between
four fields anchored to different partitions.  Each tetrahedron is
oriented so that $MB$, $PB$ and $MH$ lie in a plane.  The view 
perpendicular to that plane is shown below 
and tangent to the plane is shown
above, with the same scale.  The tetrahedron on the left is
from the partition of \fig\ \ref{fig:msk}, whose distances
appear in \fig\ \ref{fig:fount2}.  Distances are in percentages of
$\Phi_{\rm tot}$ and horizontal and vertical reference bars 
are shown ($5\%$ and $10\%$ respectively).  The center and right
columns are tetrahedra for partitions with $N=108$ and $N=395$
regions.  The vertices, labeled on the left, are in the same
orientation for all three cases.}
	\label{fig:tetra}
\end{figure}

The quantity $\Delta\Psi_{X-Y}$ can be considered the 
{\em connectivity distance}
between the coronal field models $\bvec^{(X)}$ and $\bvec^{(Y)}$.
When $\Delta\Psi_{X-Y}=0$ the fields are identical, at least with
respect to their connectivities.  For a given partition there are four
different fields separated by six distances.  The fields can be
represented as vertices of a tetrahedron in a three-dimensional 
space.  The left column of \fig\ \ref{fig:tetra} shows two views of
the tetrahedron formed from the histograms in \fig\ \ref{fig:fount2}.
The distances used in these plots have biases removed.

If all six distances were exactly the same, they would form a regular
tetrahedron.  In fact two of the distances, $\Delta\Psi_{PB-MH}$ and 
$\Delta\Psi_{MB-PH}$, are largest, leading to the extremely flat
tetrahedron shown in \fig\ \ref{fig:tetra}.  The flattened shape is
approximately a quadrilateral lying in a plane, with the two large
distances forming its diagonals.  The sides of the quadrilateral
separate vertices (fields) differing in only one respect.  The face-on
views of the quadrilateral (bottom) are oriented so that horizontal
edges separate fields of different anchoring (M {\em vs.} P)
while vertical edges separate fields of different boundaries 
(B {\em vs.} H).

Performing the same analysis for different partitions gives distances
with similar properties, as exemplified by the center and right
tetrahedra in \fig\ \ref{fig:tetra}.  Fields differing in both
anchoring and boundaries are furthest apart, so the tetrahedron is
flattened into a quadrilateral.  Furthermore both instances of a
particular difference,
such as the anchoring (the horizontal sides in the lower view), 
result in a similar distance.
This makes the quadrilateral into an approximate
parallelepiped.  For the cases with more sources (center
and right) changing boundaries makes the largest differences, so the
parallelepipeds tend to be taller than they are wide.
Finally, the similarity of the diagonals pushes the shape toward a
rectangle.

\subsection{Variation in partitions}

To see how the above comparisons are affected by different partitioning
parameters we perform Monte Carlo calculations for fields from
different partitions.  The smoothing parameter $h$ is varied from
$0.1$ to $2.0$ arcsecs, and $B_{\rm sad}$ from $10$ G to $500$ G.

\begin{figure}[htp]
\epsscale{0.8}
\plotone{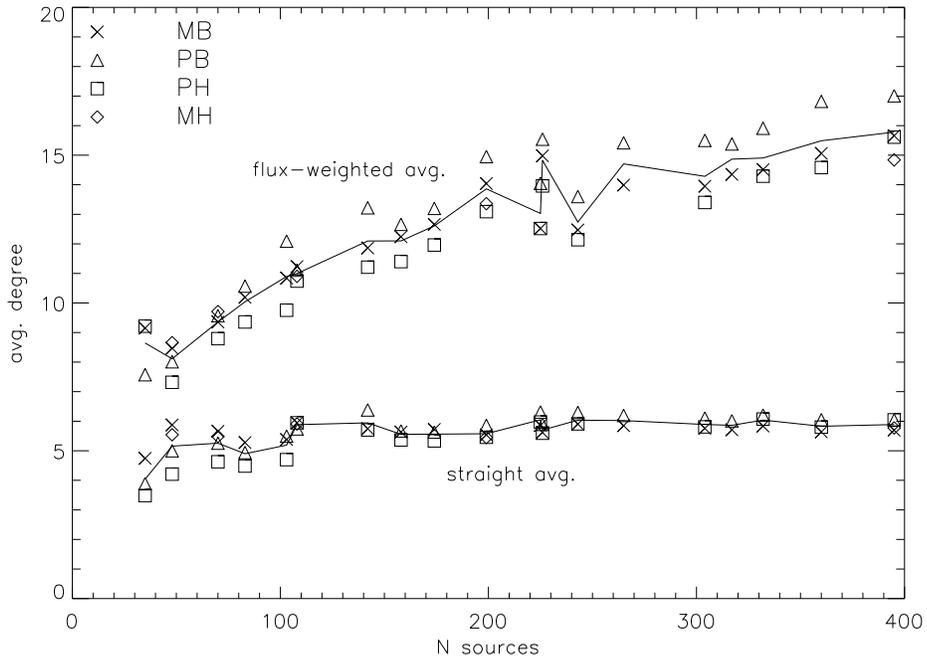}
\caption{The average degree (number of connections) per source for
various fields.  Different partitions are plotted versus
the number of sources, $N$ as symbols.  The average of all fields is
plotted as a solid curve.  The
bottom set are the average, $\bar{d}_a=2C/(N+1)$, while the
upper are the flux weighted average, $\avg{d_a}$ from \eq\
(\ref{eq:avg}).}
	\label{fig:deg_sweep}
\end{figure}

Once again we find that the different partitioning can be approximately
ordered by the number of regions, $N$.  Figure \ref{fig:deg_sweep}
shows the average degree of a source region in each of the
fields over a range of partitioning.  The lower points show the
un-weighted average, $\bar{d}_a=2C/(N+1)$, as illustrated in \fig\
\ref{fig:phi_cm_scat}.  This quantity is very similar,
$\bar{d}_a\simeq6$, for all fields and all levels of
partitioning.  It seems that the number of connections scales as
$C\sim3N$ even as the number of {\em possible} connections goes as
$\sim N^2$.  In contrast, the flux weighted average, $\avg{d_a}$,
does appear to increase with the number of regions, although perhaps at
a power less than $\sim N$.  This shows that the largest sources
connect to more sources as they become available.

Comparisons from the previous section, between all four kinds of
field, suggested that it is sufficient to compare only three.
The six possible
comparisons between all four fields were visualized as a tetrahedron
of distances.  It was found, however, that these tended to form a 
flat rectangle, well characterized by two of its sides.  
Taking advantage of this we consider the
fields $MB$, $PB$ and $PH$ for a large number of different partitions.
Among the three comparisons there is one differing only by anchoring
($MB$ versus $PB$), one differing only by boundaries ($PB$ versus
$PH$) and one differing in both respects ($MB$ versus $PH$).

\begin{figure}[htp]
\epsscale{0.8}
\plotone{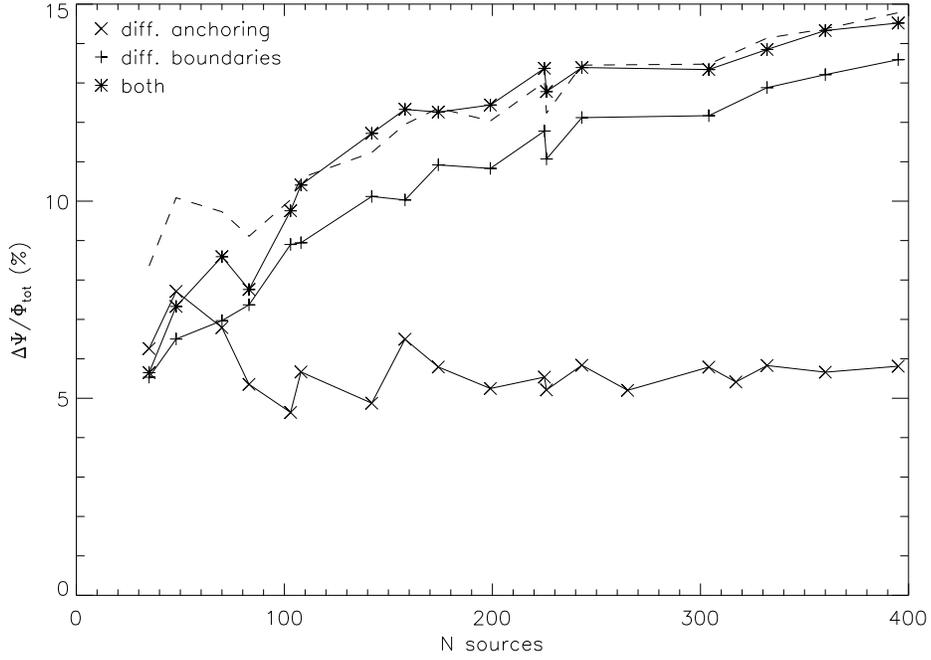}
\caption{The total connectivity difference, $\Delta\Psi_{X-Y}$,
defined in \eq\ (\ref{eq:DeltaPsi}) for different fields and different
partitions.  Comparisons are between fields with different anchoring
($MB-PB$) $\times$, different boundaries ($PH-PB$) $+$, and differing
in both ($MB-PH$) $\ast$.  
All differences are plotted as a percentage of $\Phi_{\rm tot}$.  The
dashed line shows 
$(\Delta\Psi^2_{MB-PB}+\Delta\Psi^2_{PH-PB})^{1/2}$ for comparison.}
	\label{fig:Psi_sweep}
\end{figure}

Finer partitioning (i.e.\ smaller values of $h$ or $B_{\rm sad}$)
result in more source regions, $N$.  According to the lower curve in
\fig\ \ref{fig:deg_sweep} these sources are interconnected in a
proportionately large number of ways, $C$, regardless of which coronal
field is used.  One expects that subdividing the same total flux, 
$\Phi_{\rm tot}$, into a larger number of pieces would yield bigger
discrepancies, $\Delta\Psi_{X-Y}$.  Figure \ref{fig:Psi_sweep}
shows that this expectation is borne out when comparing fields with
different boundaries, ($PB$ to $PH$, $+$s or $MB$ to $PH$, $\ast$s).  
These curves trend upward as $N$ increases.  

In keeping with the
results of the previous sub-section, the cases which differ in both
respects ($\ast$) are separated by the greatest distance.  The quantity 
$(\Delta\Psi^2_{MB-PB}+\Delta\Psi^2_{PH-PB})^{1/2}$, plotted as a
dashed line, appears to match the the asterisks well.  This fit
corroborates the observation from the previous section that the
connectivity distances formed a flat rectangle; the dashed curve is
the hypotenuse of a right triangle formed from the other two distances.

A truly remarkable feature of \fig\ \ref{fig:Psi_sweep} is that the
two bounded fields, differing only in their photospheric anchoring,
$MB$ and $PB$, differ by a relatively small amount ($\sim 6\%$) which
{\em does not} change with finer partitioning.  
This corroborates the tendency observed in \fig\ \ref{fig:tetra} for
the rectangles to grow taller with increasing $N$, without growing wider.
So while there are
ever more connections being compared, and more positive terms in \eq\
(\ref{eq:DeltaPsi}), the total difference does not seem to change.  It
seems that the difference between using point sources or using the
actual magnetogram, is about 6\% of the connectivity.  

\subsection{Variations in box size}

We can further explore the effect of outer boundaries by increasing
the size of the conducting box.  To do this the box $\bvol$ is
augmented by 
layers of equal width, $w$, along all boundaries except the bottom ($z=0$).
This new domain, called $\bvol^+$, has conducting boundaries on the
four lateral walls and a uniform field at the top boundary $z=L_z+w$.
For the field, $PB^+$, the lower boundary has the same $N$ point
sources located at the same positions within the central 
$L_x\times L_y$ square.  The field is computed on a grid with cubic
pixels, $1.1''$ on  side, just as in the $PB$ field.

\begin{figure}[ht]
\epsscale{0.7}
\plotone{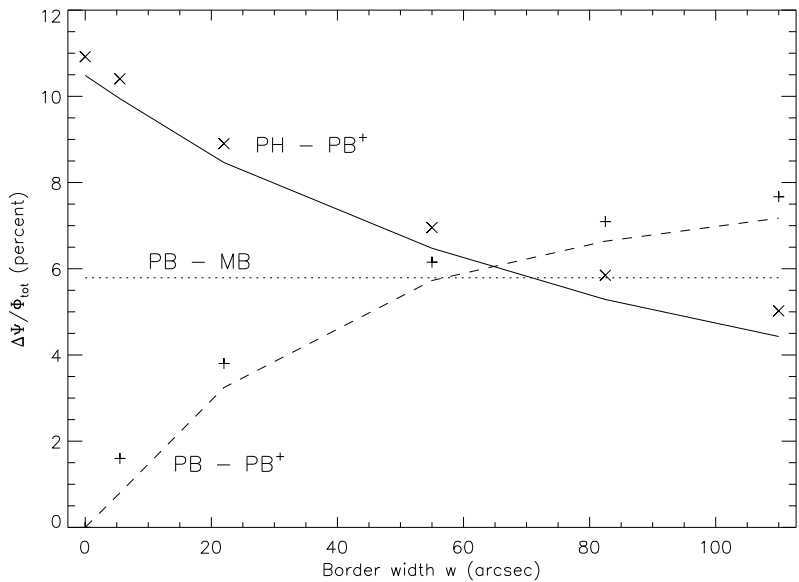}
\caption{Comparison of fields $PB^+$ to $PB$ (dashed) and $PH$
(solid) for the partition with $N=174$ regions.  
The symbols above the line are the raw differences,
$\Delta\Psi_{PB-PB^+}$ ($+$) and $\Delta\Psi_{PH-PB^+}$ ($\times$).
The corresponding lines are corrected for statistical bias.  The
dotted line shows the value $\Delta\Psi_{PB-MB}$ for comparison.}
	\label{fig:pad}
\end{figure}

Connectivities are computed in fields $PB^+$ anchored to point sources
from the partition with $N=174$ sources ($B_{\rm sat}=100$ and
$h=0.1''$).  This is computed for different boundary layer widths,
$w$, and the results are compared to the $PB$ and $PH$ fields.  For
vanishing layer width ($w=0$) the ``augmented'' volume corresponds to
$\bvol$ so $\Delta\Psi_{PB-PB^+}=0$.  In the other limit,
$w\to\infty$, $PB^+\to PH$ and $\Delta\Psi_{PH-PB^+}\to0$.  Figure
\ref{fig:pad} shows the continuous transition between these limits.
For the case with a border width $w\simeq70''$ the field $PB^+$ has
become equally dissimilar to both other fields, $PB$ and $PH$.  Borders
wider than this yield a field still closer to that of the half-space
($PH$).

\section{Discussion}

Connectivity characterizes coronal magnetic field in a manner
useful for understanding energy release and reconnection.  It is
possible to quantify the connectivity of an active region based on a
single photospheric magnetogram.  It is necessary to first construct
the coronal magnetic field using some kind of extrapolation and then
to partition the magnetogram into unipolar regions.  Techniques for
accomplishing each of these steps have been developed and are in
common use.  The foregoing work has presented new techniques for
quantifying the differences in connectivity for different
fields anchored to the same set of sources.  This comparison was used
to assess which steps the computed connectivity is most sensitive to.

Our comparisons show that the connectivity is relatively insensitive
to variations in the methods of extrapolation or photospheric
anchoring.  Among the cases we considered, the greatest discrepancy
between any two fields was 15\% 
of the total flux.  That is to say that one field may be converted
into the other, at least in terms of connectivity, by reconnecting
15\% of its field lines.  The majority of the difference occurred
in a small number of connections present in both fields but with
different fluxes.  The vast majority of connections were common to
both fields, however, there were instances (singletons) of
connections present in one field and not in the other.
These topological differences were found to occur most frequently 
in small connections  which taken together accounted for a small part of
the overall flux difference.

It is not immediately clear how large a difference one could expect 
between any two arbitrary fields anchored to the same set of
photospheric regions.  Equations 
(\ref{eq:Phib}) and (\ref{eq:Phia}) place numerous constraints on the
possible connectivities, which could render 100\% difference
impossible.  It is worth considering a few artificial connectivities for
the purpose of comparison.  One class of 
connectivities are those minimizing or maximizing the informational
entropy function,
\be
  H ~=~\sum_{a\in\regs_+}\sum_{b\in\regs_-}
  {\psi_{ab}\over\Phi_{\rm tot}}\,
  \ln\left({\Phi_{\rm tot}\over\psi_{ab}}\right) ~~,
	\label{eq:H}
\ee
subject to the constraints from \eqs\ (\ref{eq:Phib}) and
(\ref{eq:Phia}).  The partition from \fig\ \ref{fig:msk} has 46
positive regions and 25 negative regions, including $\infty$.
The informational entropy is maximized ($H=4.97$) 
by connecting the sources in all $46\times25=1150$ possible ways
($\psi_{ab}=\Phi_a|\Phi_b|/\Phi_{\rm tot}$).  It is
minimized ($H=3.07$) by a set of 66 connections.  These extreme cases
differ from one another by $\Delta\Psi=76\%$.  All versions of the
potential field extrapolations have very similar entropies,
$H\simeq3.82$ and differ from the minimum and maximum entropy
connectivities by similar amounts: $\Delta\Psi\simeq62\%$ and
$\Delta\Psi\simeq58\%$ respectively.  It seems that potential field
extrapolations are far more similar to one another then to these particular
fields.

It would be better to compare to
connectivities generated in a more realistic fashion 
than to those extremizing an {\em ad hoc} function.  We could ask, for
example, how different is the connectivity of a potential field from
that of a NLFFF extrapolated from the vector magnetogram; or we could
compare the potential field from one time to that at another time
(provided the magnetograms are partitioned into equivalent regions).
Comparisons of this kind promise insight into energetics and
reconnection in real coronal fields, and will be the topic of future
investigation.  In order to gain this physical insight, however,
it is essential to know the level of difference that
arises from non-physical variations such as in
anchoring or boundary conditions alone.  The present study
provides that important point of reference, and will therefore serve
as a baseline in the future studies.

Connectivity difference has proven itself a useful metric for
quantifying discrepancy between different 
coronal extrapolations from the same data.  
The presence or absence of conducting
boundaries are found to have the greatest effect on the
connectivities of a potential field.  Figure \ref{fig:pad}
corroborates the expectation that more distant boundaries give
a better approximation of no boundaries at all.
The $260''\times221''$ magnetogram considered here can be
expanded to four times the area, by padding with $w=120''$ on
all sides, to produce a field only 4\% different from that in an
infinite half-space.  It is possible that proportionately more padding would
be required for magnetograms with better flux balance, since these
would have longer closed loops.  Expression (\ref{eq:r0}), giving the
extent of the closed field, is inversely proportional to the degree of
balance.  A future study will seek a general expression
for the required padding to a given magnetogram.

Alternatively, the field from an infinite half-space field, $MH$, can
be computed on a grid after using a Green's function to compute it on
all lateral boundaries.  Unless this grid is large enough, 
$\sim r_0$ from \eq\ (\ref{eq:r0}), there
will be connections extending outside the grid which cannot be
followed precisely.  It might be possible to follow them approximately 
with a gridless field, such as $PH$, in a kind of hybrid method.
Alternatively, it appears that the point source approximation alone,
$PH$, is a relatively accurate approximation to $MH$ (differing by 
roughly $5.5\%$ in our case) for which a grid is not
necessary.

We find that only a small error (5\%--6\%) is incurred when
connectivity is computed using 
a simplified, gridless extrapolation from magnetic point charges
($PH$ or $PB$) in place of more traditional extrapolation from
a full magnetogram ($MB$ or $MH$).  
These point-charge models differ significantly 
from the actual field: for example they are singular at
the charges.  The connectivity, however, seems only mildly 
sensitive to these local differences.  Moreover the connectivity
difference does not increase even as the number of
source regions, and therefore the number of connections, increases.
This seems explicable by the fact that the point source anchoring
differs from the magnetogram only within a small neighborhood of the
charge.  The differences may therefore be confined to a layer 
$z\la \avg{r_g}$ which shrinks with finer partitioning.

Large scale connectivity is defined in terms of unipolar source
regions into which the photospheric field (magnetogram) 
is partitioned.  Variation of
parameters controlling this partitioning leads to significant 
changes in the source regions and therefore 
the connectivity.  At least for the two parameters whose variation we
explored, $h$ and $B_{\rm sad}$, most differences could be ordered just
by the number of regions $N$.  The sizes, shapes and
interrelation of regions appears to scale with $N$, as did most
properties of the potential field connectivity.

Smaller values of partitioning parameters $h$ or $B_{\rm sad}$ led to
finer partitioning, with more sources and therefore more
connections.  Remarkably we found that the total number of
connections increased only as $N$ rather than as $N^2$ like the number
of {\em possible} connections.  Indeed, we found that this
particular active region had approximately 6 connections to every
source independent of partitioning parameters.  Further study will
reveal whether this trend persists in other active regions.

\acknowledgements

The work was supported by a grant from NASA's Living with a Star TR\&T
program.  GB was partially 
supported for this work by the Air Force Office of
Scientific Research under contract FA9550-06-C-0019.

\appendix

\section{Estimating and correcting bias in the absolute value}

Consider an unknown quantity $x$ whose measurement, $\tilde{x}$,
includes an additive Gaussian error
of known variance $\sigma^2$.
The absolute value of the measurement, $|\tilde{x}|$, is an estimate of
$|x|$ whose expectation is
\be
  \avg{|\tilde{x}|} ~=~ \int_{-\infty}^{\infty}
  |x+\epsilon|\,p(\epsilon)\, d\epsilon ~=~
  |x| ~+~ \sqrt{2\over\pi}\sigma e^{-x^2/2\sigma^2} ~-~
  |x|\,{\rm erfc}(|x|/\sqrt{2}\sigma) ~~,
	\label{eq:abs_err}
\ee
where erfc is the complementary error function.  The second and third
terms on the right represent a bias in the estimate of $|x|$,
\be
  E_b(x) ~=~ \sqrt{2\over\pi}\sigma e^{-x^2/2\sigma^2} ~-~
  |x|\,{\rm erfc}(|x|/\sqrt{2}\sigma) ~~,
	\label{eq:bias}
\ee
since its expectation does not vanish.  
For values $|x|\gg\sigma$ the bias error is 
extremely small ($\sim \sigma e^{-x^2/2\sigma^2}$), 
and $\avg{|\tilde{x}|}\simeq|x|$.  For small magnitudes, on the other
hand ($|x|\ll\sigma$), the estimate will be
dominated by the magnitude of the Gaussian noise so 
$\avg{|\tilde{x}|}\sim 0.8\,\sigma$.

The actual bias, $E_b$, depends on the quantity $|x|$ whose value we are
trying to learn from the measurement $\tilde{x}$.  We cannot,
therefore, subtract the exact bias from the measurement.
We must instead construct a function of the measured value, $\tilde{x}$, 
whose expectation
approximates $E_b(x)$.  This function has a
discontinuous derivative at $x=0$, due to its second term, and is
therefore very
difficult to reproduce in the expectation of a function of
$\tilde{x}$.  The expectation of a given function $f(\tilde{x})$
can be expressed as the convolution of $f$ with the Gaussian distribution
of $\epsilon$.  This convolution effectively blurs $f(x)$ over a scale
$\sigma$, thereby smoothing out discontinuities.

Because of the blurring property described above subtracting
$E_b(\tilde{x})$ would remove a broader function from the
expectation of the estimate.  We seek instead a function more sharply
peaked, whose convolution will be limited to $|x|\la\sigma$.
Following this logic we propose the function
\be
  B(\tilde{x},\sigma) ~\equiv~ {\sigma\alpha\over\sqrt{2\pi}}\,
  e^{-(\alpha\tilde{x}/\sigma)^2/2} ~~,
	\label{eq:bias_est}
\ee
where $\alpha$ is an adjustable parameter defining the width.  The
expectation of the function
\be
  \avg{B} ~=~{\sigma\over\sqrt{2\pi}}\,
  {\alpha\over\sqrt{\alpha^2+1}}\,
  \exp\left[-{\alpha^2x^2\over 2\sigma^2(\alpha^2+1)}\right] ~~,
	\label{eq:bias_exp}
\ee
resembles the first term in
(\ref{eq:bias}) and has the same integral as the actual error
\be
  \int_{-\infty}^{\infty} \avg{B}\, dx ~=~
  \int_{-\infty}^{\infty} E_b(x)\, dx ~=~ \sigma^2 ~~,
	\label{eq:bias_int}
\ee
independent of $\alpha$.

The bias estimator in \eq\ (\ref{eq:bias_est}) is limited to
$|\tilde{x}|\la\sigma/\alpha$, and 
in the limit $\alpha\to\infty$ it
becomes a Dirac $\delta$-function: 
$B(\tilde{x})\to\sigma^2\delta(\tilde{x})$.  It is natural that in the 
$\delta$-function limit the expectation, (\ref{eq:bias_exp}), is
simply the distribution of noise.
For large $\alpha$ the bias correction 
will only rarely be non-negligible; then it will be
large to compensate for the numerous times it was negligible.  
This will introduce additional variance to inferred
values.  Adopting instead a small value of $\alpha$ will subtract
a small amount from more measurements, owing to the much broader scope
of $\avg{B}$.  We have found $\alpha=3$ to be a
reasonable all-around compromise since its scope is very narrow while
introducing little additional variance.

\begin{figure}[htp]
\plottwo{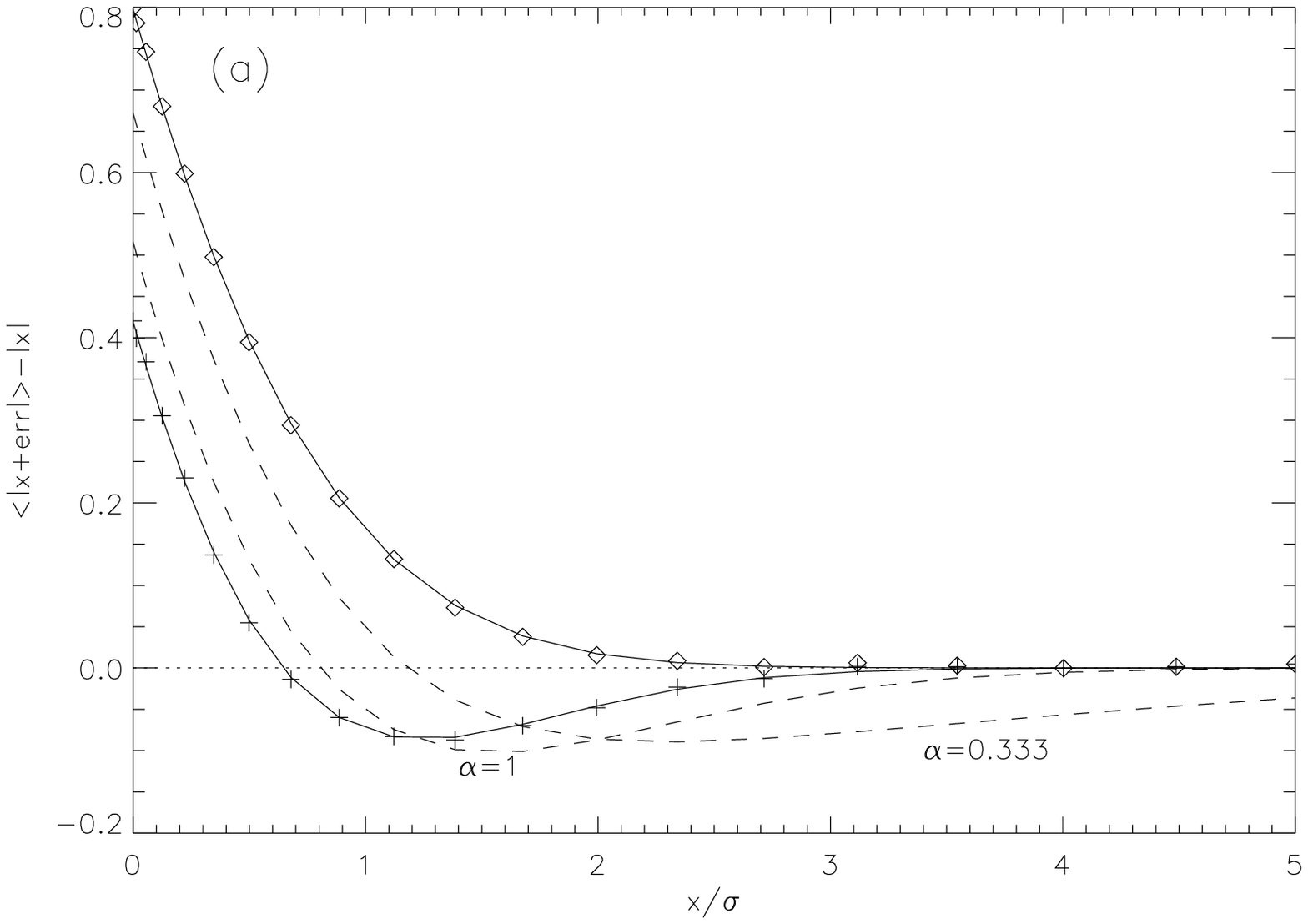}{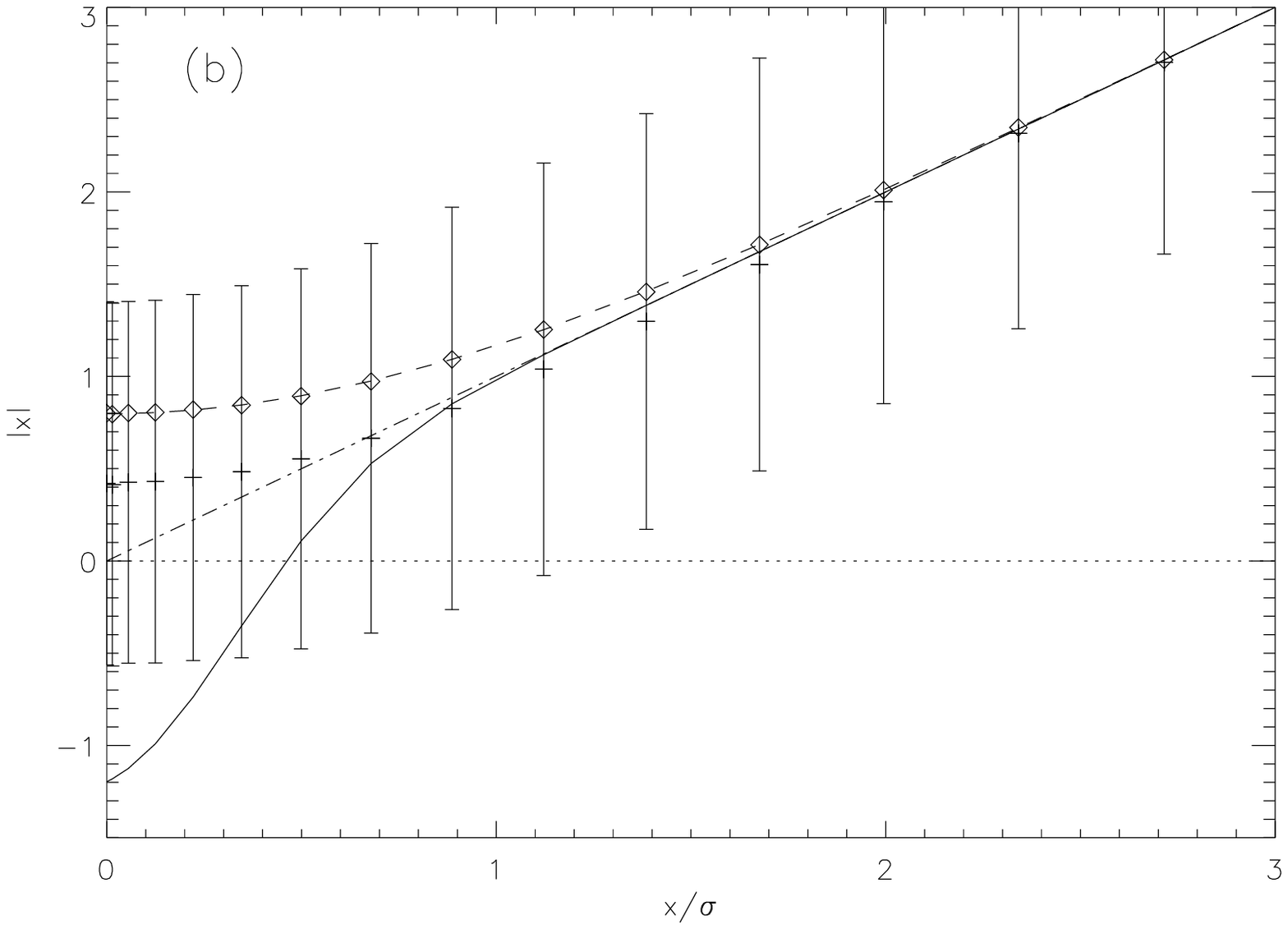}
\caption{Effects of the bias estimator $B(\tilde{x})$ on estimates of
$|x|$.  (a) The bias error as a function of $|x|$ (in units of
$\sigma$).  Diamonds show $E_b(x)$, the bias in $\avg{|\tilde{x}|}$.
Dashed lines show $E_b\avg{B}$ for choices
$\alpha=1/3$ and $\alpha=1$.  The pluses are for the value used in
practice: $\alpha=3$.  Symbols are the results of $10^5$ trials
with Gaussian noise, while the lines are the analytical functions from
\eqs\ (\ref{eq:bias}) and (\ref{eq:bias_est}).  (b) The computed
values with and without compensation.  Diamonds and plusses are, as in
(a), the results of averaging $|\tilde{x}|$ and
$|\tilde{x}|-B(\tilde{x})$ respectively.  Error bars are the standard
deviation in the $10^5$ trials.  The solid line shows the
actual value $|x|-B(x)$ versus $|x|$.}
	\label{fig:bias_comp}
\end{figure}

The problem we face is to compute a sum of magnitudes of
measurements, $\tilde{x}_i$, of different underlying values, $x_i$.
We estimate this by the sum
\be
  \sum_i|x_i| ~\simeq~ 
  \sum_i\Bigl[\, |\tilde{x}_i| - B(\tilde{x}_i,\sigma_i)\,\Bigr] ~~,
\ee
where $B(\tilde{x},\sigma)$ 
is defined by \eq\ (\ref{eq:bias_est}) with $\alpha=3$.
Terms of the sum where $|\tilde{x}_i|<0.46\sigma_i$ are negative
in order to remove the bias.  These negative terms, as well as 
positive terms where $|\tilde{x}|<0.7\sigma$, actually
underestimate the bias on average (see \fig\ \ref{fig:bias_comp}). 
In order to compensate, those values in the range
$0.7\sigma<|\tilde{x}|\la 2.5\sigma$ over-estimate it on average.
Provided the underlying values, $x_i$, are distributed
relatively uniformly within the range $|x|\la 2.5\sigma$, the
underestimates and overestimates will balance one another, due to \eq\
(\ref{eq:bias_int}), thereby removing the bias precisely.  Even when
this is not the case, the bias error at an individual value of $x$ is
reduced by at least half.

%\bibliography{/home/dana/tex/inputs/short_abbrevs.bib,/home/dana/tex/full_lib.bib}
\bibliography{/Users/danalongcope/stuff/short_abbrevs.bib,/Users/danalongcope/stuff/full_lib.bib,local.bib}

\end{document}